\documentclass{JHEP3}
\usepackage{graphicx}
\usepackage{epsfig,amsmath,amsthm}
\newcommand{\be}{\begin{equation}}
\newcommand{\ee}{\end{equation}}
\newcommand{\bea}{\begin{eqnarray}}
\newcommand{\eea}{\end{eqnarray}}
\def\bse{\begin{subequations}}
\def\ese{\end{subequations}}
\newcommand{\IR}{\mathbb{R}} 

\def\IZ{\relax\ifmmode\hbox{Z\kern-.4em Z}\else{Z\kern-.4em Z}\fi}
\newcommand{\IS}{\mathbb{S}} \newcommand{\IT}{{\bf T}}

\newcommand{\non}{\nonumber \\}

\def\half{\frac{1}{2}} 

\def\del{{\partial}} 
\def\room{~\rule[-2mm]{0mm}{8mm}}

 \def\co{{\cal O}}
  
\def\cL{{\cal L}} 
  \def\eps{\epsilon}

\def\lam{\lambda} 
\def\htau{{\hat \tau}}

\def\room{~\rule[-2mm]{0mm}{8mm}}
\def\presub{\vspace{.5cm} \noindent}

\def\bi{\begin{itemize}} \def\ei{\end{itemize}}

\def\Schw{Schwarzschild }
\def\({\left(} \def\){\right)}
\def\[{\left[} \def\]{\right]}
\title{ \center{Classical Effective Field Theory and Caged black holes}}

\author{Barak Kol and Michael Smolkin\\
Racah Institute of Physics, Hebrew University\\
Jerusalem 91904, Israel\\
E-mail:
{\tt\href{mailto:barak_kol@phys.huji.ac.il}{barak\_kol@phys.huji.ac.il}},
\email{smolkinm@phys.huji.ac.il}}

\abstract{Matched asymptotic expansion is a useful technique in
General Relativity and other fields whenever interaction takes
place between physics at two different length scales. Here matched
asymptotic expansion is argued to be equivalent quite generally to
Classical Effective Field Theory (ClEFT) where one (or more) of
the zones is replaced by an effective theory whose terms are
organized in order of increasing irrelevancy, as demonstrated by
Goldberger and Rothstein in a certain gravitational context.
 The ClEFT perspective has advantages as the procedure is clearer,
it allows a representation via Feynman diagrams, and divergences
can be regularized and renormalized in standard field theoretic
methods.
 As a side product we obtain a wide class of classical examples of
regularization and renormalization, concepts which are usually
associated with Quantum Field Theories.

We demonstrate these ideas
through the thermodynamics of caged black holes, both simplifying
the non-rotating case, and computing the rotating case.
 In particular we are able to replace the computation of six
two-loop diagrams by a single factorizable two-loop diagram, as
well as compute certain new three-loop diagrams. The results
generalize to arbitrary compactification manifolds.
 For caged rotating black holes we obtain the leading correction
for all thermodynamic quantities. The angular momentum is found to
non-renormalize at leading order.}

\begin{document}

\section{Introduction and Summary}

Matched Asymptotic Expansion (MAE) is an analytical tool which
applies to problems containing two (or more) separate scales. In
mathematical physics this idea goes back as far as Laplace who
used it to find the shape of a drop of liquid on a surface - see
\cite{Damour} and references therein for a historical review. In
\cite{dialogue1,dialogue2} Gorbonos and one of the authors (BK)
applied MAE to the problem of small caged black holes, namely
black holes which are much smaller than their background
compactification manifold. On p.7 of \cite{dialogue2} it was
recognized that the divergences which appear at higher order of
MAE and their regularization are ``reminiscent of renormalization
in Quantum Field Theory''. In \cite{CGR} Chu, Goldberger and
Rothstein applied an effective field theory approach, rather than
MAE, to the same problem of small black holes, thereby simplifying
the derivation of their thermodynamics and extending it to a
higher order. That work built on the ideas of Goldberger and
Rothstein regarding an effective field theory of gravity for
extended objects \cite{GoldbergerRothstein1}; see also
\cite{GoldbergerRothstein2} and a pedagogical introduction in
\cite{Goldberger-Lect}.

In this paper we further develop these ideas. The paper is
composed of three parts. In the first part, subsection
\ref{MAE-to-ClEFT}, we argue for a quite general equivalence of
MAE and effective field theory. In the other parts we proceed to
apply and illustrate these ideas in the context of the
thermodynamics of caged black holes. In the second part, we start
in subsections
\ref{set-up-subsection}-\ref{subsection-non-quantum} by describing
several improvements to the effective field theory analysis of
caged black holes, and we continue in section \ref{caged-section}
to significantly economize the derivation of the thermodynamics of
static caged black holes, and to perform a new computation.
  %3-loop
In the third part, section \ref{rotating-section}, we apply the
method to obtain new results for rotating caged black holes. We
end this introduction with a summary of results.

\subsection*{General Equivalence of MAE and EFT}

In subsection \ref{MAE-to-ClEFT} we claim that Matched Asymptotic
Expansion (MAE) is equivalent quite generally to an effective
field theory. We observe the phenomena of regularization and
renormalization in this classical set-up and as a way of stressing
it we refer to the method as Classical Effective Field Theory
(ClEFT). Even though ClEFT is formally equivalent to MAE we
indicate the advantages of the ClEFT perspective: a clear
representation via Feynman diagrams, the effective action as a way
of studying a zone once and for all, easy power counting, and the
use of dimensional regularization. Finally we characterize quite
generally the domain of validity of ClEFT to be whenever an
extended object (such as a soliton) moves in a background whose
length scale is much larger than the object's size.

\subsection*{Caged black holes}

As a concrete realization of the ideas regarding the equivalence
of Matched Asymptotic Expansion and Effective Field Theory we apply
 them to the problem of caged black holes, namely black
holes in the background $\IR^{\hat{d}} \times X$ where the
compactification manifold will be taken to be $X=\IS^1$ throughout
most of the paper.

This problem was motivated by the effort to establish the phase
diagram of the black-hole black-string transition
\cite{TopChange}, see the reviews \cite{review,HOrev} and
references therein. The problem was engaged with a combination of
analytic and numeric methods. It was studied analytically in
\cite{H4} using adapted coordinates in a single patch; in
\cite{dialogue1} with a two-zone MAE; in \cite{KSSW1} the
asymptotic thermodynamics properties were computed to
$\co\(m_0^3\)$ in 5d MAE; in \cite{dialogue2} the $\co\(m_0^2\)$
correction was found for all $d$ together with a systematic
discussion of regularization; Finally in \cite{CGR} effective
field theory was used to compute to order $\co\(m_0^3\)$ for all
$d$. Numerical studies include a 5d simulation \cite{KPS}; a 6d
simulation \cite{KudohWiseman6d} relying on an earlier brane-world
simulation \cite{KudohTanakaNakamura} and finally
\cite{KudohWiseman5d} which produced significantly larger 5d black
holes. For another perspective see  a review of
``phenomenological'' work on black holes in theories with large
extra dimensions \cite{KantiRev}.

In section \ref{ingred-section} (except for subsection
\ref{MAE-to-ClEFT}) we study static caged black holes. We present
several improvements to the ClEFT method which allow us to
reproduce the results of Chu, Goldberger and Rothstein \cite{CGR}
(and to perform a new computation).
%3-loop
 Actually we believe that we
have finally \emph{discovered the shortest route to these
results}.  The main improvements to the method are  \bi
 \item We perform a change of variables through a dimensional reduction
over $t$. It has the advantage
that the propagator is diagonal with respect to the field $\phi \simeq h_{00}/2$
which couples to the world-line at lowest order (through the
interaction $(-)m_0\, \phi$ ).
 \item It is shown that the mass renormalization $\delta m$ can be
read off a zero-point function, rather than a 1-point function.
Moreover, this zero-point function serves as a partition function
and thermodynamic potential. From it we are able to derive the
tension, temperature and area (completing all equations of state)
at the price of computing another quantity, the red-shift, up to a
similar order.
 \item We note several points where the classical nature of our
problem allows simplification through the elimination of certain
quantum features which appear in the approach of Rothstein and
Goldberger \cite{GoldbergerRothstein1,Goldberger-Lect} which is
based on a background in Quantum Field Theory (QFT). These
features are: Planck's constant $\hbar$ (implicit in the
definition of the Planck mass), the complex number $i$ and Feynman
path integrals.
 \ei

Through these improvements we obtain the following results in
section \ref{caged-section} \bi
 \item \emph{Our method replaces the 6 diagrams required
for the 2-loop computation of $\delta m$ in \cite{CGR}
(fig.\ref{sixdiag}) by a single diagram
(fig.\ref{Feynman_diag_Sch}(b))!} Moreover, it does not require the
quartic vertex of GR, nor the cubic one as it happens. This
diagram happens to manifestly factorize and hence is simple to
compute, thereby explaining the factorization which was observed
in \cite{CGR}.

 \item We reproduce other thermodynamic quantities: the
 temperature (\ref{Sch_temperature}), tension and entropy
 (\ref{Sch_tension_entropy}).

 \item Ignoring finite-size effects it is now possible
to proceed to higher orders. We perform part of
%3-loop
the calculation  of
$\delta m$ to order $\co\(m_0^4\)$ (\ref{m-caged-3-loop}).
Neglecting finite-size effects is justified for $d < 7$, which
happen to be the numerically studied cases.

 \item We point out that the results immediately generalize
to any (Ricci-flat) compactification manifold.
 \ei

\subsection*{Rotating black holes}

In addition to economizing the computations for static black holes
(and performing a certain extension thereof), we apply in section
\ref{rotating-section} the ClEFT method to compute for the first
time the caging effect on the thermodynamics of the rotating
Myers-Perry black holes \cite{MyersPerry}.

The thermodynamics of caged rotating black holes could be useful
for determining a black-hole black-string phase diagram in the
presence of angular momentum. Non-uniform rotating black string
solutions were studied in the case of equal angular momenta in 6d
\cite{KKR}. In the rotating case the isometry group is much
reduced and accordingly a MAE analysis would require a much larger
number of metric components. In ClEFT thermodynamics, on the other
hand, the reduced symmetry hardly manifests itself and up to the
relevant order all we need to add are several new world-line
vertices. Actually determining these vertices is one of our motivations
 as a step towards the full effective action of moving and spinning black holes.

We proceed to describe our main results. We determine the two
leading world-line vertices which involve the local angular
momentum $j_0$ (figure \ref{Feynman_rules_Kerr}) and confirm that
they agree with the existing literature
\cite{ReggeHanson,Porto:2006bt,Porto:2005ac}. We compute the
leading expressions for all thermodynamic quantities: mass
(\ref{m-rot}), angular momentum (\ref{ang_moment}), temperature
and angular velocity (\ref{Kerr_temperature}), and finally tension
and area (\ref{rot-tau-j-s}). In the computation of the mass we
proceed to compute $\delta m$ to order $\co\(j_0^2\) \simeq
\co\(m_0^2\, r_0^2\)$. Somewhat unexpectedly we find that although
the mass renormalizes \emph{the angular momentum does not
renormalize} at leading order (\ref{ang_moment}). It would be
interesting to know whether this non-renormalization holds to all
orders and if so to prove it. It could be especially interesting
if insight from the mechanism behind this non-renormalization
could apply to non-renormalization in QFT as well.

{\bf Note added (v2)}. Minor changes upon publication. These
include: a global change in sign $J \to -J$  to conform with
standard conventions, appearing in
(\ref{Kerr_eff_action},\ref{Gamma_eff_Kerr}, \ref{4.9}) and fig.
\ref{Feynman_rules_Kerr} which does not affect the final results;
a factor of $2$ in (\ref{vector_prop});
 and at the top of p.3 in the first item we corrected that our change of variables
should not be considered a change of gauge.

\presub \emph{Address.} In presenting this paper we hope that it
would be especially enjoyed by Gorbonos, Chu,  Goldberger and
Rothstein upon whose work we build here.

\section{Main Ingredients}
\label{ingred-section}

We start with a general discussion of the equivalence of Matched
Asymptotic Expansion (MAE) with Classical Effective Field Theory
(ClEFT). Then we proceed to set-up the problem of caged black
holes, and discuss several more specific improvements to the
method in that case.

 \subsection{From MAE to ClEFT}
 \label{MAE-to-ClEFT}

Matched Asymptotic Expansion (MAE) entails the use of two zones
(or more) at widely separated scales. In each zone one of the scales is fixed while the other
is infinitely small or infinitely large. The interaction dialogue
between the scales occurs through supplying each other with
boundary conditions. For instance, applying MAE to the problem of
caged black holes \cite{dialogue1,dialogue2} requires two zones:
the near zone where the black hole has fixed size $r_0$ but the
compactification scale is infinitely far, and an asymptotic zone
where the compactification size $L$ is fixed and the black hole is
point-like and fixed to the origin.

\emph{Divergences} and the associated need for \emph{regularization} were observed to appear at higher orders in the small parameter $r_0/L$ \cite{dialogue2}. The first instance was while
solving for the next to leading correction in the asymptotic zone.
In that zone the leading correction is simply the Newtonian
potential of the point-like object, which solves a Laplace-like
equation. At the next order one needs to solve a similar equation,
only the non-linear nature of General Relativity (GR) introduces a
source term quadratic in the Newtonian potential (and its
derivatives). Since the Newtonian potential diverges near the
origin (the location of the object) this source term has an even
worse divergence, and the Green's function integral diverges.
Quantitatively, the first correction to the metric $h^{(1)}$ is
determined by the Newtonian potential $\Phi$ which solves
$\triangle \Phi \propto \delta(\vec{r})$; it behaves as $\Phi \sim
1/r^{d-3}$ where $r$ is the distance from the black hole, and $d$
is the total space-time dimension; the equation for the second
order perturbation to the metric $h^{(2)}$ is schematically
$\triangle h^{(2)} = Src \sim (\del \Phi)^2 \sim 1/r^{2(d-2)}$;
hence the Green function integral behaves as $h^{(2)}(x') = \int
dx\, G(x',x)\, Src(x) \sim \int_{\eps} r^{d-2} dr\, /r^{2(d-2)}
\sim 1/\eps^{d-3}$, and certainly diverges for all relevant
dimensions ($d \ge 5$).

The concept of \emph{renormalization} can also be seen to arise
in the  context of caged black holes. The simplest example is
the mass of the black hole:  while an observer at a distance
$r_0 \ll r \ll L$ measures the local mass $m_0$, a distant asymptotic
observer at $r \gg L$ measures a different mass $m$, which is
slightly smaller, the leading effect being the Newtonian binding
energy between the black hole and its images.
This can be interpreted as a dependence of the mass on the length scale at which it is measured,
exactly in the spirit of renormalization.

Recall that historically, divergences quite similar to these
obstructed the development of Quantum Field Theory (QFT) for about
two decades from soon after the discovery of quantum mechanics in
1926-7 till their treatment with counter-terms and the completion
of the theory of Quantum Electro-Dynamics (QED) in 1948. It took
even longer, till the early 1970's to reveal the renormalization
significance of regularization. Actually several Nobel prizes were
awarded for these achievements: to Feynman, Schwinger and Tomonaga
for QED and to Wilson for the theory of second order phase
transitions which is intimately connected with renormalization.

It is quite obvious that at the time physicists were not familiar
with any examples of regularization and renormalization,
definitely not in classical physics. Even today we are not
familiar with too many such examples (the authors would appreciate
correspondence on this issue). % Omit in journal version
A notable exception is the classical regularization and renormalization near the boundary of
Anti-de-Sitter space \cite{AdS-renorm}.

Here we claim that \emph{Matched Asymptotic Expansion is
equivalent quite generally to an effective field theory}.  The
equivalence is achieved by replacing one (or more) of the zones by
a point-like effective action (usually it is the near zone but we
may also consider replacing an asymptotic zone by effective
boundary conditions at the asymptotic region of the near zone).
The physics of the eliminated zone is coded in various interaction
terms in the effective action. A more precise statement of the
equivalence is that the ClEFT is equivalent not to all the
observables of the MAE but rather to those which do not reside in
the replaced zone.

Let us present a general argument for this claim. Whenever we have
two (or more) widely separated length scales, we may cleanly
decompose the fields into corresponding components by performing a
spatial Fourier transform and dividing the field according to the
scale of the spatial frequencies. This is equivalent to the
decomposition into zones. Then we may integrate out the field
component in one of the zones. Integrating out a near-zone with
its high spatial frequencies is analogous to a Born-Oppenheimer
approximation which integrates out fast degrees of freedom. The
definition of classically integrating out a field and its
eligibility are discussed in subsection
\ref{subsection-non-quantum}. This integration replaces \emph{by
definition} the integrated field or zone with interaction terms in
the effective action. The opposite direction poses an interesting
question, namely to what extent can the action in the
integrated-out zone be reconstructed given the effective action.

The concepts of regularization and renormalization appear in these
equivalent methods quite generally. We stress the fact that this
happens in a completely classical set-up by referring to the
theory as a \emph{Classical Effective Field Theory, acronym
ClEFT}. This perspective provides a large class of new
classical examples of renormalization and regularization (some
features will appear only in non-linear theories), a class which
includes boundary layer phenomena in hydrodynamics, waves in a
background with defects and black holes moving in a slowly varying
background. These examples may be useful in building classical insight
into the appearance of the same concepts in QFT.

Despite ClEFT being equivalent to MAE it does offer several
advantages in perspective, as well as in practical computations
\bi
 \item Feynman diagrams provide as a clear representation of
 the computation.

 \item The near zone needs to be considered only once.

In MAE one alternates between zones. In ClEFT we need to go once
through the process of replacing the near zone by an effective
action (at least up to a prescribed order), and then we can forget
 the near zone altogether.

 \item Easy power counting.

In ClEFT the effective interaction terms are ordered by powers of
the small parameter which determine an order of relevancy.
Accordingly the power counting of each Feynman diagram is easily
recognized in terms of the vertices which appear in it.

 \item Dimensional regularization.

While in GR one of the standard regulators is Hadamard's (Partie
finie), which requires certain care and attention to case by case details,
 the field theory perspective suggests dimensional
regularization which proves to be symmetry preserving,
straight-forward and efficient in ClEFT just as it is in more
standard QFT.
We find dimensional regularization to be equivalent to the regularization
used in \cite{dialogue2} -- it is equivalent to Hadamard's method
as both are essentially an analytic continuation, and it is seen
to realize the no-self-interaction feature (see subsection \ref{m-subsection}).

\ei Some of these points appeared already in \cite{CGR}.

We proceed to stress and clarify two general points which
appear in \cite{GoldbergerRothstein1}.

The first issue involves the domain of validity. The central
application of \cite{GoldbergerRothstein1} is to the
Post-Newtonian expansion for the radiation from an inspiraling
binary. In that expansion the small parameter is the velocity $v
\ll 1$. However, the natural domain of ClEFT is wider and simple
to state: \emph{ClEFT is valid whenever an extended object (such
as a soliton) moves in a background whose length scale is much
larger than the object's size}. Similar statements for the
gravitational context appear in \cite{GoldbergerRothstein2}. Note that ClEFT applies
not only to gravity and would be equally useful to say
 a monopole placed in a non-trivial background
in a Yang-Mills theory. Returning to the binary
inspiral problem the system contains two independent dimensionless
parameters: the velocity $v$ and the mass ratio $m_1/m_2$. ClEFT
applies not only when $v \ll 1$ but also when $m_1/m_2 \ll 1$
and while the first condition always fails at the last stage of the inspiral,
the second condition allows in principle to
compute the radiation throughout the whole evolution in a controlled
way.

The second issue is the classical nature of problem. While
\cite{GoldbergerRothstein1} is rooted in a QFT background the
problem at hand is classical and as such it allows for certain
adjustments in the theory (more precisely certain quantum issues
can be avoided and left out of the theory). The ingredients which
can be avoided include Planck's constant $\hbar$, the complex
number $i$ and the Feynman path integral, as we discuss in detail
in subsection \ref{subsection-non-quantum}.

\subsection{Caged black holes  -- set-up}
 \label{set-up-subsection}

For concreteness we turn to consider static caged black holes, and we start by setting-up
the problem. Consider a compactification background of the
form $\IR^{\hat{d}} \times X$ where $X$ is a compact manifold. For
simplicity we take the theory to be pure gravity (though
additional fields could be accommodated) and hence $X$ is assumed
to be Ricci-flat. The total space-time dimension is
$d=\hat{d}+{\rm dim} X$. We make another simplifying assumption by
considering mostly  $X=\IS^1$, a circle of size $L$ parameterized by the coordinate $z$
(see however subsection \ref{arbit-subsection}), and accordingly
$d=\hat{d}+1$.

Next we consider placing a small static black hole at a point in
$\IR^{\hat{d}} \times X$. As long as a certain no-self-force is
obeyed the black hole will remain at rest. Here we do not need the
explicit form of the no-force condition and it suffices to observe
that certain symmetries are enough to guarantee it. If  the
black hole position $p \in X$ is a fixed point of an isometry\footnote{More
 generally an isolated member of the fixed-set of a subgroup of isometries.}
then the force vector must vanish. Moreover, assuming there is at least one
equilibrium point in $X$ (this must be true because there is no external
energy source), then if $X$ is homogeneous any point in it would
be an equilibrium point. Since $X=\IS^1$ is both homogeneous and
enjoys the discrete symmetry of inversion ($z \to 2 z_0-z$ for arbitrary $z_0$), any
point on $\IS^1$ is an equilibrium point.

Our aim is to compute the thermodynamics of this system, as
encoded by the fundamental thermodynamic relation $G=G(\beta,\Omega_i,L)$
where $\beta$ is the inverse temperature of the black hole and $\Omega_i$
are the angular velocities in the rotating case.

The basic feature of the problem of small caged black holes is
that we have two widely separated length scales \be
 r_0  \ll L \label{scales} \ee
where $r_0$ is the \Schw radius. Accordingly the metric (and any other field) can be decomposed \be
 g_f \supset  g \supset \bar{g} \ee
 where $g_f$ is the full metric including all length scales,
 $g$ includes only length scales of order $L$ or
larger, and
finally  $\bar{g}$ includes
only length scales much larger than $L$ and can be thought to live
at the asymptotic region. We sometimes write $g_f=g_S + g$ where
$g_S$ is the component of the metric field with short length
scales of order $r_0$, and $g=g_L+\bar{g}$ where $g_L$ is the $L$-scale component of the metric.

The original action is purely gravitational, without any source
terms \be
 S = \int R[g_f] ~.\ee
 Our basic tool is to integrate out the short degrees of freedom around the
black hole and replace them by an effective world-line action \be
 S_{eff}[g] = I[S,g_S]=\int R[g] + S_{BH}[g,x,e^\mu_I] \ee
 where we denote by $I[S,g_S]$ ``integrating $g_S$ out of $S$'' as
defined below in subsection \ref{subsection-non-quantum}.  The black hole effective action depends on $x,\, e^\mu_I,\, g$. The first two are black hole
degrees of freedom: $x=x(\tau)$ is its location while ``the frame'' $e^\mu_I=e^\mu_I(\tau)$ is
 a rotational degree of freedom. $g$ represents here the local background at the location of the black hole.

The black hole effective action, $S_{BH}$, needs to be evaluated only once (up to the required order)
and then it can be used to study BH motion through any background (whose typical length scale is much larger than the black hole). Naturally $S_{BH}$ must be invariant
under word-line reparameterization as well as the more general
background diffeomorphisms. Its leading term is the point-particle
 action $S_p$ characterized by the (local) mass $m_0$  \bea
 S_{BH} &=& S_p + \dots \non
 S_p &=& -m_0\, \int d\tau \label{pp-action1} \eea
 where $d\tau \equiv \sqrt{g_{\mu\nu}(x)\, dx^\mu\, dx^\nu}$ is the
proper time interval along the world-line, and the ellipsis denote terms which depend on gradients of the background.

\subsection{Dimensional reduction and the Newtonian potential}

Caged black holes have a time translation symmetry which we now
turn to exploit. Given this symmetry there exists a natural change
of variables, namely the outcome of a \emph{Dimensional reduction}
over $t$. The new variables will be especially useful to simplify
the computations since $ g_{00}$ which appears in the leading
(mass) term of the world-line effective action (\ref{pp-action1})
will be separated from the other metric components and mapped onto
a scalar $\phi$.

Dimensional reduction is commonly used to reduce over a compact spatial dimension which the
fields do not depend upon, but it can be used equally well for
reducing over the non-compact temporal direction, as long as all
the fields are $t$-independent. The standard Kaluza-Klein ansatz
is given by \be
 ds^2 = g_{\mu\nu}\, dx^\mu\, dx^\nu
      = e^{2 \phi}\, \(dt-A_i\, dx^i\)^2 - e^{-2\phi/(d-3)}\, \gamma_{ij}\, dx^i dx^j \label{KKansatz} ~,\ee
which defines a change of variables $g_{\mu\nu} \to
(\gamma_{ij},A_i,\phi)$. We let Greek indices run over all
coordinates while Latin indices are spatial, namely $\mu \to (t
\equiv 0,i)$. Note that our signature convention for $g$ is mostly
minus, $(+-\dots-)$, as in field theory, while for the purely
spatial metric $\gamma_{ij}$ we change the signature to be all $(+)$. In particular the scalar
field $\phi$ is defined through \be
 e^{2\phi}=g_{00} \label{def-phi} ~.\ee

Since in the stationary case  the action is proportional to $\int dt$ we may factor
it out and define a reduced action \be
 S_R:=S\left/ \int dt \right. \label{reduced-action} \ee
 where from hereon we shall suppress the subscript `R'.
The resulting bulk action is \bea
 S &=& \frac{1}{16\pi G}\int R[g] \rightarrow \non
 \rightarrow S &=& -\frac{1}{16\pi G} \int dx^{d-1} \sqrt{\gamma}
  \[ R + \frac{d-2}{d-3} \(\del \phi\)^2 - \frac{1}{4}\, e^{2(d-2)\phi/(d-3)} F^2 \]~,\label{bulk_action} \eea
 where the second line displays the reduced action (\ref{reduced-action}) and in which only the metric $\gamma$ is being used: $R=R[\gamma], ~~\(\del \phi\)^2 = \gamma^{ij}\, \del_i \phi\,
\del_j \phi$ including the standard definitions $F^2=F_{ij} F^{ij},
 ~~F_{ij}=\del_i A_j- \del_j A_i$.
 The action describes a metric $\gamma_{ij}$ with a standard
Einstein-Hilbert action (this is achieved through the Weyl rescaling
factor in front of $\gamma_{ij}$ in the ansatz) a negative-kinetic-term vector field $A_i$ with
 a $\phi$ dependent
pre-factor, and a minimally coupled scalar field $\phi$ which is
related to $g_{00}$. The negative kinetic term for $A_i$ is directly related to the fact that the spin-spin force in gravity has an opposite sign relative to electro-dynamics, namely ``north poles attract'' \cite{Wald-spin}. Finally the constant pre-factor $(d-2)/(d-3)$ which appears in the kinetic term for $\phi$ is related to the polarization dependence of the $g$ propagator (the original graviton), see footnote\footnotemark[7].
%\footnotemark (d-2)/(d-3)

Given the time-translation symmetry we can be more specific regarding the
black hole effective action (\ref{pp-action1}). The BH stands at a spatial point which we denote
as the origin $O$. The BH degrees of freedom are frozen
(more precisely the velocity $\dot{x}$ and the angular velocity $\Omega^\mu_\nu$ are frozen)
 and  $S_{BH}=S_{BH}[g]$.
 This action must be supplemented by certain no-force and no-torque constraints,\footnote{In our applications these constraints will be satisfied automatically due to symmetry. See the discussion of the no-force constraint in the previous subsection.}
 which originate from the equations of motion for $x,\, e^\mu_I$.
 The (reduced) point-particle action becomes
 \be S_p=-m_0\,  \sqrt{g_{00}} = -m_0\, e^\phi \label{Sch_world_line_action} \ee
 where all field are to be evaluated at $O$ and in the second equality we used the change of variables
(\ref{def-phi}).

There is a nice physical interpretation for the scalar field
$\phi$ which appears in the dimensional reduction. It is a free
scalar field which couples to the mass and as such it is quite similar to the Newtonian potential in Newtonian gravity.
Its equation of motion
to leading order (first in $\phi$ and zeroth in the other fields)
is \be
 \triangle \phi = 8 \pi G\, \frac{d-3}{d-2}\, m_0\, \delta(x) \label{phiEOM-lin}\ee
where $\triangle$ is the flat-space spatial Laplacian.

In our system there are additional symmetries beyond time
translation, namely azimuthal symmetries. In principle one could
dimensionally reduce over the corresponding angles, but in this
paper we reduce only over time which is special both because the
associated scalar field $\phi$ appears in the leading world-line
interaction and because a reduction over time is more generic and
applies to all stationary sources (neutron stars, ordinary starts
etc.) which unlike black holes do not necessarily possess an azimuthal
symmetry. Actually in higher dimensions the Myers-Perry black
holes are symmetric with respect to $[(d-1)/2]$ angles
while so far a proof guarantees a single angular symmetry for a general higher-dimensional black hole \cite{HIW}.

\subsection{Vacuum diagrams}

\begin{figure} [t!] \centering \noindent
\includegraphics[width=9cm] {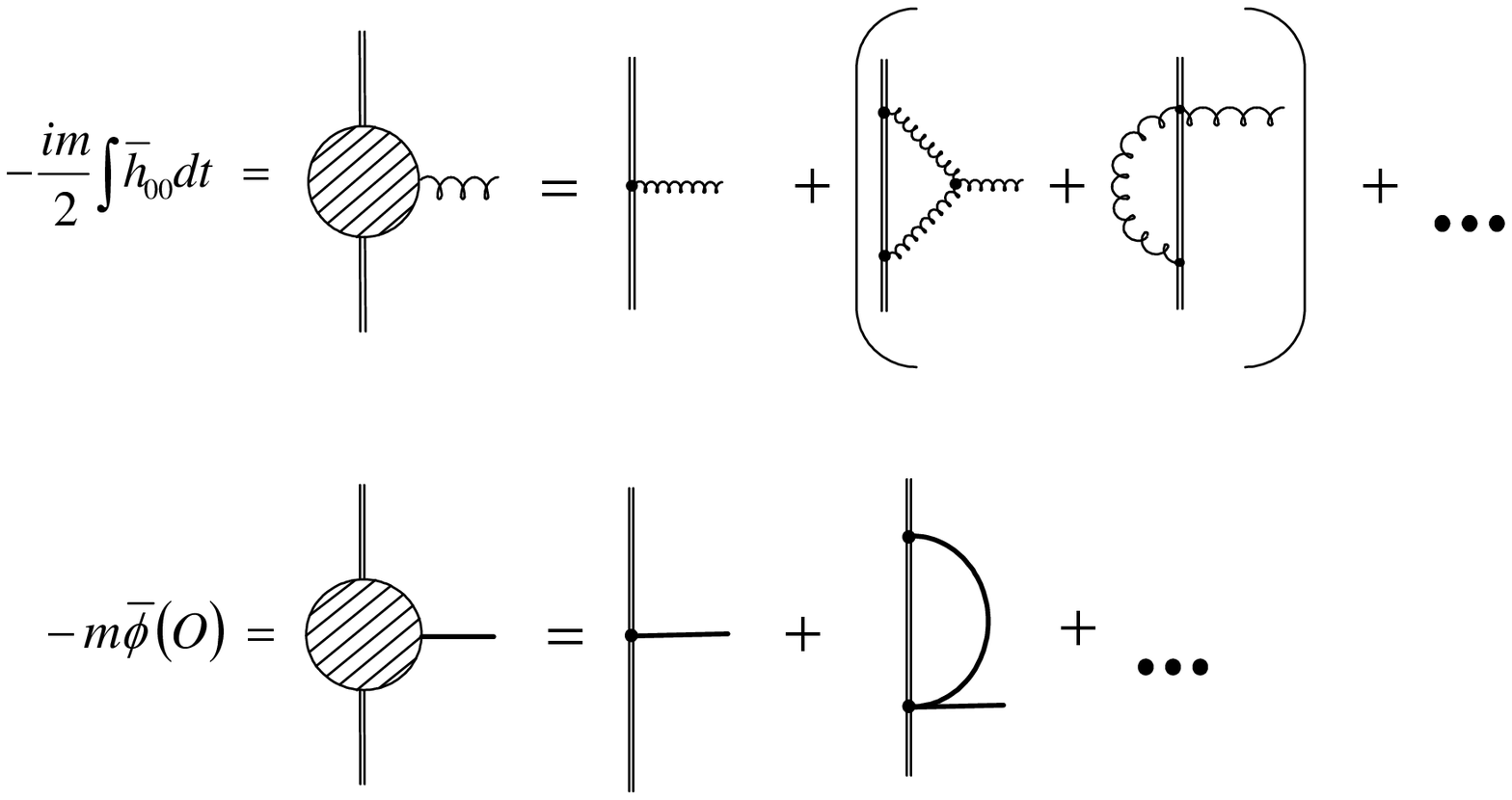}
\caption[]{A definition of the (asymptotic, renormalized) ADM mass
$m$ in terms of a 1-point function as in \cite{CGR}. The top line is pre-dimensional reduction
as in \cite{CGR} and the wavy lines represent gravitational perturbations (of $g$). The external leg is the asymptotic component $\bar{h}_{00}$ and there is no propagator associated with it. The
double solid lines denote the black hole world-line. The bottom line
 is the translation of the top line into the dimensionally reduced fields
which we use. $\bar{\phi}$ is the
longest scale ($\gg L$) component of the metric field related to
$g_{00}$ through \ref{def-phi}. The solid internal lines denote the
propagator (\ref{scalar_prop}) for the scalar field $\phi$.  More details
about the Feynman rules will be given later in figure
\ref{Feynman_rules_Sch}.} \label{Feynman_diag_Sch3}
\end{figure}

A central objective is to calculate the (asymptotic) ADM mass $m$
given $m_0$, the local mass of the caged black hole. As remarked
already this can be considered to be a renormalization of the mass
from the scale $r_0 \ll r \ll L$ where $m_0$ is defined to the
scale $r \gg L$ where $m$ is defined.  In \cite{CGR}  $m$ was
calculated from a 1-pt function relation, which when translated to
the language of dimensional reduction, is represented by the
Feynman diagrams in figure \ref{Feynman_diag_Sch3}.

\begin{figure} [t!] \centering \noindent
\includegraphics[width=9cm] {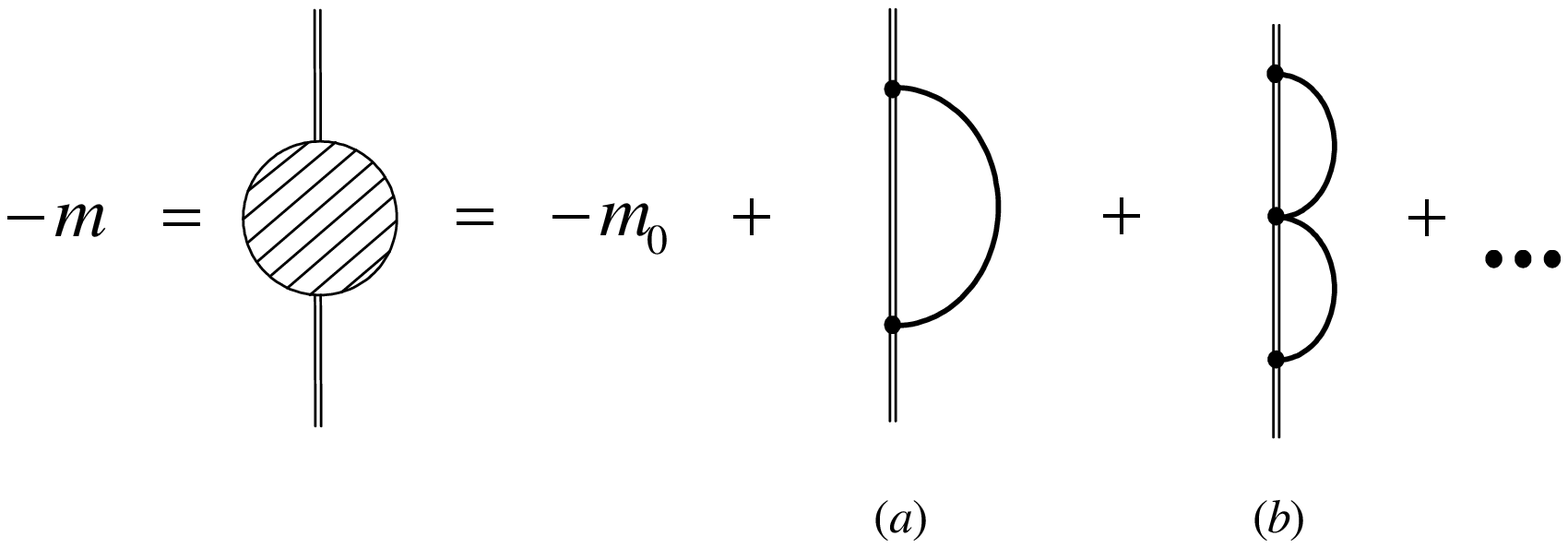}
\caption[]{The new definition of $m$ as a 0-point or ``vacuum''
function, represented in terms of Feynman diagrams. (a) and (b)
represent correction to the mass $m$ of different orders (1 and 2
respectively). The order parameter $\lambda$ (related to $r_0/L$)
will be defined later in (\ref{def-lambda}). More details
about the Feynman rules will be given later in figure
\ref{Feynman_rules_Sch}.}
\label{Feynman_diag_Sch}
\end{figure}

We suggest an improved calculational definition for $m$
represented by the Feynman diagrams in figure
\ref{Feynman_diag_Sch}.
 This definition avoids the need for an external leg and along with it
reduces the maximum ``connectivity index'' of the required
vertices (for example, unlike \cite{CGR} we avoid using the 4-graviton vertex
at two loop, as well as the 3-vertex, as it happens).

We explain fig.\ref{Feynman_diag_Sch} first intuitively and then
more formally. Intuitively, the classical vacuum diagrams describe
the vacuum energy at the location of the black hole, namely its
mass $m$, while $m_0$ is the tree level value. Technically one
needs to prove that the two definitions are identical. To see that
we integrate out the scale $L$ component of the metric $g_L$. By
residual diffeo invariance the action must look like
$S_{eff}[\bar{g}]=-\int d\tau\, m$ where the proper time element
is $d\tau=\sqrt{\bar{g}_{\mu\nu} dX^\mu
dX^\nu}=\sqrt{\bar{g}_{00}}\, dt=e^{\bar{\phi}}\,
dt=(1+\bar{\phi}+\dots)\, dt$ (or equivalently
$d\tau=(1+\bar{h}_{00}/2 + \dots)\, dt$) and $m$ is a scalar.
 Thus the coefficient of the two terms in the
effective action must coincide and we can indeed identify the two
diagrammatic definitions
(fig.\ref{Feynman_diag_Sch3},fig.\ref{Feynman_diag_Sch}) of the mass $m$.

 \subsection{The thermodynamic potential}
 \label{thermo}

In the previous subsection we defined the black hole mass in terms
of ``vacuum'' diagrams \be m=m(m_0,a_i,L) \ee
 where $m$ is considered as a
function of the local black hole parameters, namely the local mass $m_0$
and the local rotation parameters $a_i$ (in higher dimensions
there are several independent rotation parameters), and the
parameters of the compactification such as $L$. As such it is
analogous to the ``vacuum energy'' or ``partition function'' of
translationally invariant quantum field theories, which is known
to encode all properties of the QFT (more precisely, the partition
function in the presence of arbitrary sources encodes in its
derivatives all the correlation functions of the QFT). On the
other hand in thermodynamics, and in particular in black hole
thermodynamics, all the thermodynamic information is known to be
contained in a ``fundamental thermodynamic relation'' specifying a
thermodynamic potential in terms of its natural variables. In this
subsection we will build from the mass function such a
thermodynamic relation.

It is convenient to choose horizon parameters as our basic
variables, namely the temperature $T$ (or equivalently the surface
gravity) and the angular velocities $\Omega_i$. Correspondingly
the relevant ensemble is the grand canonical one and the potential
is the Gibbs potential \be
 G =G (T,\Omega_i,L) ~.\ee
 From derivatives of $G$ one can infer all the equations of state,
namely the expressions for the entropy, tension and angular
momenta $S=-\del G/\del T,~\htau=\del G/\del L,~J_i=-\del G/\del
\Omega_i$ \footnote{We denote the tension by $\htau$ to prevent
any possible confusion with the proper time $\tau$.}.
 In the static (non-rotating case) where $\Omega_i$ do not appear the
ensemble coincides with the canonical ensemble and the potential
becomes the free energy.

The natural parameters which appear in the computations are
$m_0,a_i,\,L$ while the natural parameters for $G$ are $T,\Omega_i,\,L$
 and therefore we need to find the relation between these
two sets of parameters. This is done through an intermediate set
$T_0,\Omega_0$, the local temperature and angular velocities (we
omit the index $i$ from the angular velocities for clarity of
notation) which are defined to be the quantities measured by an
``intermediate'' observer at a distance $r_0 \ll r \ll L$ from the
black hole. Schematically the transformation is \be
 (m_0,a) \to (T_0,\Omega_0) \to (T,\Omega) \ee
 The first transformation is carried out considering the black
hole to be embedded in Minkowski space-time, namely by using the
standard relations of the Myers-Perry black hole
\cite{MyersPerry}.

\begin{figure}[t!]
\centering \noindent
\includegraphics[width=9cm]{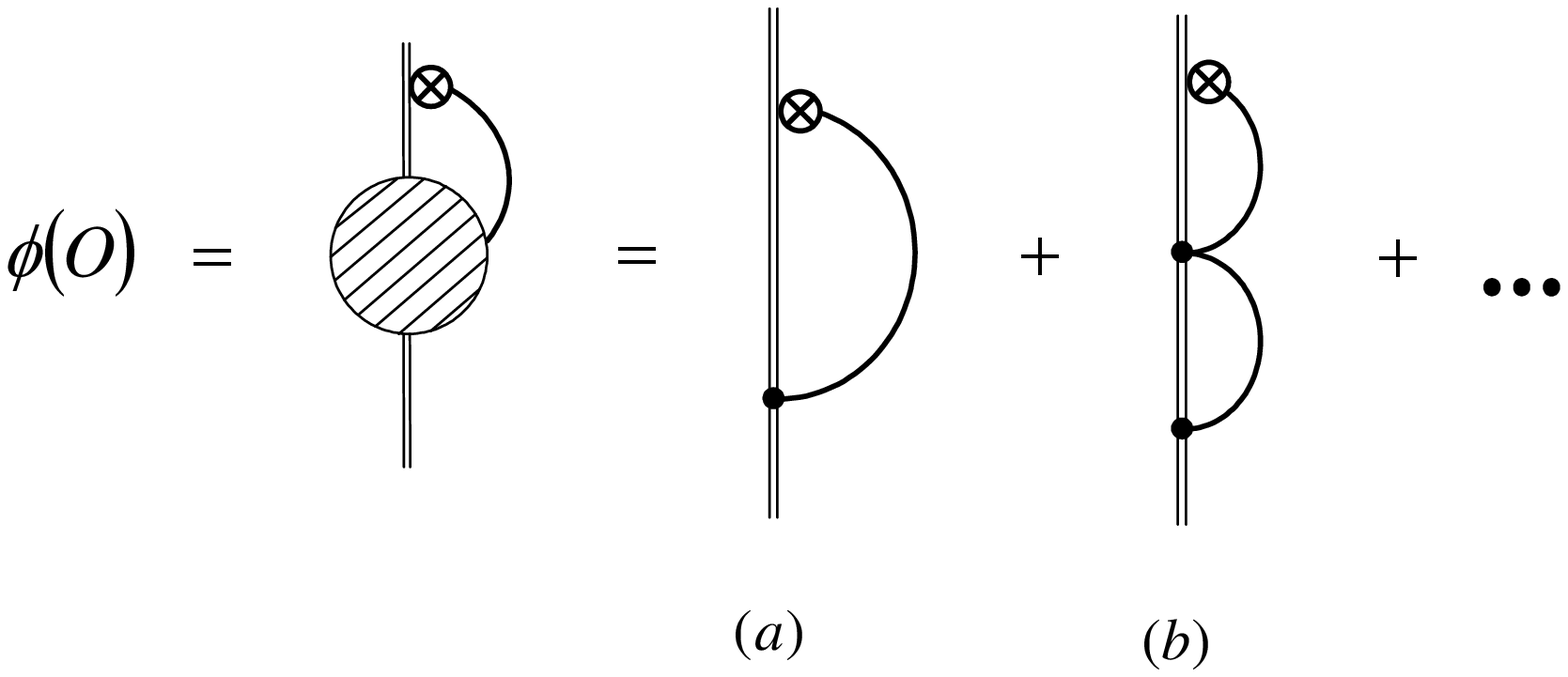}
\caption[]{A Feynman diagram definition of the red-shift factor
$R$.
 (a) and (b) denote contributions to  different orders (1 and
2) in the order parameter $\lambda$ which will be defined later in
\ref{def-lambda}. The $\otimes$ denotes the value of $\phi$ at the
location of the black hole, that is $\phi(O)$.} \label{g00}
\end{figure}

In order to perform the second transformation, from local horizon
quantities to their values for an asymptotic observer another
ingredient is needed. While the intermediate and asymptotic
observer agree on their definition of proper distances, their
notion of time differs due to the red-shift factor \be
 R:=\sqrt{g_{00}(O)} \equiv e^{\phi(O)} = \frac{t_0}{t} \label{red-shift}\ee
 where $g_{00}(O)$ is the metric at the black hole
location (after the scale $r_0$ component of the metric was
integrated out), and $t,\, t_0$ are the asymptotic and intermediate
times. In terms of the red-shift the asymptotic angular velocity
is given by $\Omega=dl/dt=R\, dl/dt_0=R\, \Omega_0$, where $dl$ is
an element of proper distance on the horizon (more formally one
should obtain the angular velocity from the coefficients in the
decomposition of the Killing generator of the horizon into the
time translation and angular shift Killing vectors). The
asymptotic temperature (and surface gravity) is canonically
conjugate to time and hence transforms inversely to $t$, namely
$T/T_0=t_0/t=R$ (alternatively, the Hawking temperature is
red-shifted exactly according to $R$).
 Altogether the asymptotic quantities
are given by \bea
 T &=& R\, T_0 \non
 \Omega &=& R\, \Omega_0 ~.\eea
The red-shift itself \be R=R(m_0,a_i,L,\dots) \ee may be defined
in terms of Feynman diagrams as in fig.\ref{g00}.

We note that our formulae for the transformation $(m_0,a_i) \to
(T,\Omega_i)$ are stated and were tested up to the order which we
need, but not necessarily beyond.

Having found the relation between the natural parameters of the
problem and the natural parameters for thermodynamics we need to
transform the mass function into the Gibbs potential. They are
related through the standard thermodynamical relation \be
 G:= m-T\, S - \Omega\, J ~. \label{Gibbs}\ee
More details on method of computation will be given in the next
sections.

We can now compare our method with that of CGR \cite{CGR}. There
the mass and tension are computed separately in terms of 1-point
functions. Additional thermodynamic variables of interest such as
the angular momentum would need to be computed separately as well.
In the current method on the other hand we compute a single
thermodynamic potential $m$ and it has no external legs thereby
requiring simpler vertices and simplifying the computation. The
price to pay is the computation of the red-shift (which is roughly
of the same computational difficulty as $m$) and the introduction
of possibly lengthy thermodynamic manipulations.

Another difference is that while in CGR $m_0,a_i$ are interpreted
as bare couplings which have no physical meaning here they are
assigned meaning by relating them to horizon quantities.

 \subsection{Non-quantum field theory}
\label{subsection-non-quantum}

In this subsection we stress several issues where the
\emph{classical} nature of our problem allows simplification in
comparison with quantum field theory (QFT). The approach of
Rothstein and Goldberger
\cite{GoldbergerRothstein1,Goldberger-Lect} is based on a
background in QFT and as such factors of $\hbar$ (implicit in the
definition of the Planck mass) and the complex number $i$ are
commonplace as well as Feynman path integrals. However, from the
classical point of view it is clear that all factors of $\hbar$
must cancel and hence should not appear in the first place.
Similarly, since the action and all quantities are real there is
no reason for $i$'s to appear. Finally Feynman path integrals in
QFT are notoriously difficult to define rigorously despite their
long-time usage. On the other hand in classical physics we do not
have any such apparent uncertainty in the definition of the theory
and accordingly we would expect to be able to do away with this
notion in a classical set-up. Here we believe we clarify these
points, which are perhaps not very deep, but still quite useful.

The discussion of $\hbar$ factors is related to the issue of units
in classical GR vs. QFT, which we proceed to review. In classical
physics we have three fundamental dimensions time, length and mass
denoted $T,\,L$ and $M$ respectively. Special relativity
introduces the speed of light $c$ as a universal constant, and
setting it to 1 identifies $T \equiv L$, leaving us with two
fundamental dimensions $L,M$ which continue to be the fundamental
dimensions in classical field theory (ClFT). In QFT we introduce
$\hbar$ as a second fundamental constant. Setting it to 1 allows
us to identify $L \equiv 1/M$, and it is standard custom to
measure all QFT quantities in units of mass. In GR on the other
hand $\hbar$ is absent but rather $G$ is introduced as a second
fundamental constant identifying $M \equiv L^{d-3}$ and it is
useful to keep $L$ as the fundamental dimension. Accordingly the
dimensions of various field theory quantities vary between ClFT,
QFT and GR, see table \ref{table1}. Note especially that in GR the
Lagrangian density has dimensions $1/L^2$, canonical scalar fields
are dimensionless and the elements of the Fourier space dual to
space-time is better described as wavenumbers rather than momenta.
The notion of the Planck mass merits another comment. In quantum
gravity we have both $G,\hbar$ and hence we have a natural unit
for each dimension. Up to numerical constants of convention one
defines the Planck length $l_P^{d-2}= G\, \hbar$ and accordingly
the Planck mass $m_P=\hbar/l_P=\hbar^{(d-3)/(d-2)}\,/G^{1/(d-2)}$.
Since $m_P$ requires $\hbar$ for its definition it has no place in
a classical theory.

\begin{table}[t!]
\centering $\begin{array}{l|c c c}
        & ClFT  & QFT   & GR \\ \hline
 \room \mbox{action} & M\, L & 1     & ~~L^{d-2}~ \\
 \mbox{Lagrangian denity} &
          M/L^{d-1} & M^{d} & 1/L^2 \\
 \mbox{wavenumber} & L^{-1} & M & L^{-1} \\
 \mbox{canonical scalar field}~ & ~~\(M/L^{d-3}\)^{\frac{1}{2}}~ & ~~M^{(d-2)/2}~ & 1 \\
\end{array}$
\caption[]{Dimensions of various physical quantities compared
 between three types of field theories: classical (ClFT), quantum (QFT)
 and General Relativity (GR). $M$ denotes the mass dimension, and $L$ length.} \label{table1}
\end{table}

\begin{figure} [t!] \centering \noindent
\includegraphics[width=5cm]{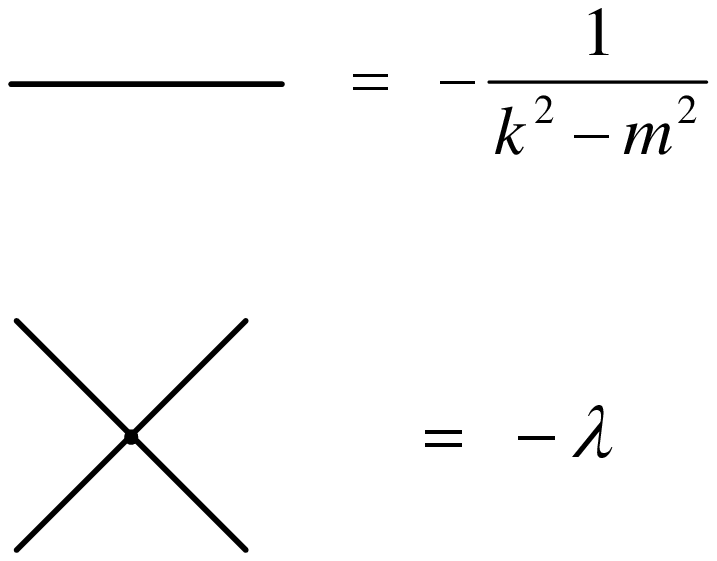}
\caption[]{The Feynman rules for ClEFT are naturally real. As an
example we display them for the $\phi^4$ scalar theory.}
\label{phi4}
\end{figure}

A second inheritance from QFT is to include factors of $i$ in the
Feynman rules for every vertex and propagator. The reason for that
is that in the functional integration (sum over histories) the
weight factor is $\exp{iS}$. In ClFT on the other hand
all quantities are real and there is no reason to have any $i$'s
in the formulae. Indeed by reviewing the origin of the Feynman
rules (for the computation of the effective action or otherwise)
one finds that all factors of $i$ can be omitted:
interaction terms come with the same sign as in the action, while
the propagator gets an additional minus sign multiplying the inverse of the kinetic term (from ``moving it to
the other side of the equation''). For example given the Lagrangian
of the $\phi^4$ theory $\cL=(\del \phi)^2\, /2 - (m^2\, \phi^2 /2 +
\lambda\, \phi^4 /4!)=- \phi\, (\del^2+m^2) \phi\, /2 - \lambda\,
\phi^4 /4!)$ the Feynman rules are given by fig.\ref{phi4}, namely
the 4-vertex is given by $(-\lambda)$ while the propagator is
$+1/(\del^2+m^2)=-1/(k^2-m^2)$. Actually from this discussion it is
clear that factors of $i$ can be omitted not only from ClFT but
also from tree-level QFT computations (at least of $S_{eff}$).
One could be motivated to generalize this to all QFT
computations. When loops are added an amendment is required which
is seen (by comparison with the standard rules) to be a factor of
$(-i)$ for each (bosonic) loop. As usual a fermionic loop adds another
negative sign so its contribution would be\footnote{One may wonder whether
additional phases are required under some circumstances in order to agree with the standard
prescription. In all cases which we checked this is not necessary, and the key point is the assumption
that $V-P+L=1$ where $V,\, P$ and $L$ are the number of vertices, propagators and loops, respectively, in the diagram. This relation holds for Lorentz invariant field theories and may hold even more generally.} $(+i)$.

Our last issue concerns the Feynman path integrals which are a
standard tool in QFT. In particular they are used in the
definition of an effective action resulting from ``integrating
out'' a field, for example \be
 i S_{eff}[g]/\hbar = \int D g_S \, \exp(i S[g_S,g]/\hbar) \label{FPI} \ee
 where $g_S,\, g$ are the short and long wavelength components of
a certain metric $g$. However, as intuitive and as useful as path
integrals are, they are also notoriously difficult to define
rigorously. In classical physics there clearly should not be any
reason for such uncertainties nor for the
loop diagrams which appear in (\ref{FPI}) at higher orders of
$\hbar$. So we would expect there exists a purely classical
definition of the effective action. Indeed it is known that in the
classical limit the path integral reduces in the saddle point
approximation to a computation around a classical solution.
Accordingly we should define \emph{``integrating out'' a field $g_S$
classically} out of the action $S$, denoted $I[S,g_S]$, as follows \be
  S_{eff}[g] \equiv I[S,g_S] := S[g, g_S(g)]
 \label{Cl-int-out} \ee
where the right hand side means that one should first solve for
$g_S$ given the prescribed long wavelength $g$ and then evaluate
the action. This definition is natural and does not produce any
uncertainty. Moreover, it stresses that \emph{one is allowed to
integrate out only when the remaining fields can specify a
solution}, for instance, fields on the boundary (or asymptotic
fields in the case of an unbounded space).

\section{Caged black hole: improved calculation}
 \label{caged-section}

\subsection{Action and Feynman rules}
 \label{Action&Feynman_rules}

Let us consider an isolated static black hole in a background with
a single compact dimension--- $\IR^{d-2,1} \times \IS^1$, where
$d$ is the total space-time dimension. Coordinates on $\IR^{d-2,1}$
are denoted by $(x^0,{\bf x})$ and $z$ labels the coordinate along
$\IS^1$. The asymptotic period of the $\IS^1$ is $L$ (as
measured by an observer at $|{\bf x}|\rightarrow\infty$).
In addition the black hole is static, thus one can take
$x^\mu=(t,{\bf x}=0,z=0)$ without loss of generality.

As a first step towards the action and the corresponding
derivation of Feynman rules  we integrate out the short degrees of
freedom $g_S$ and replace the space-time in the vicinity of the
horizon with an effective Lagrangian for the black hole world-line
coupled to gravity. The resulting effective action takes the form\footnote{Due to the
symmetry $t\rightarrow -t$ the vector field $A_i$ vanishes in the
static case.} (see
(\ref{reduced-action}-\ref{Sch_world_line_action})) \bea
 S_{eff}(g,\phi) &=& -\frac{1}{16\pi G} \int dx^{d-1}
\sqrt{\gamma}\[ R[\gamma] + \frac{d-2}{d-3} \(\del \phi\)^2 \]
\non
 &-& m_0 e^{\phi(O)} + \dots ~,
\label{eff-Lagrangian}\eea
 where the ellipsis denote finite-size higher order terms which we shall not require.
 Such terms depend on the values of the fields at the origin and respect diffeomorphism and
world-line reparameterization invariance. The leading finite-size
term is\footnote{Note that the operator $\del_i \phi(O)\del^i
\phi(O)$ is redundant (can be removed by field re-definitions, see for instance \cite{Goldberger-Lect})
since $\del_i \phi=0$ for a stationary black hole, whereas operators
involving the Ricci tensor are redundant since the black hole is placed in a Ricci flat
background.} \be
 \co := \del_i \del_j \phi(O)\, \del^i \del^j \phi(O) ~. \label{def-co} \ee

Starting from the effective action (\ref{eff-Lagrangian}) we decompose the
metric tensor and the scalar field into a long wavelength
non-dynamical background fields $\bar \phi,\, \bar{\gamma}_{ij}$
which live at the asymptotics and the scale $L$ fields
$\phi_L,\, \gamma_{Lij}$ \bea
 \phi &=& \phi_L+\bar{\phi}  \non
  \gamma_{ij}&=&\gamma_{Lij}+\bar{\gamma}_{ij} \label{decompose} ~.\eea
Since we assume the perturbative regime (\ref{scales}) all the fields are weak and
therefore it is consistent to linearize about flat space \be
 \gamma_{ij}=\eta_{ij} + \delta \gamma_{ij} \ee

Integrating out in the sense of (\ref{Cl-int-out}) or equivalently
(\ref{FPI}) the scale $L$ fields $\gamma_{Lij}$, $\phi_L$ while
holding the black hole world-line fixed leads to an effective
action $\Gamma_{eff}[\bar{\phi},\overline{\delta\gamma}_{ij}]$
valid on a scale much larger than $L$. According to
(fig.\ref{Feynman_diag_Sch3},fig.\ref{Feynman_diag_Sch}) the relation
between the ADM mass $m$ for a caged black hole and the local mass
$m_0$ can be read off from either the constant or the linear term
in $\Gamma_{eff}[\bar{\phi},\overline{\delta\gamma}_{ij}]$ \be
 \Gamma_{eff}[\bar{\phi},\overline{\delta\gamma}_{ij}] = -m-m\bar\phi(O)+\dots \label{Gamma_eff} \ee
  The constant term is represented by the Feynman diagrams of
fig.\ref{Feynman_diag_Sch},
whereas the linear term is represented by the Feynman diagrams of
fig.\ref{Feynman_diag_Sch3}.

\begin{figure}[t!]
\centering \noindent
\includegraphics[width=7cm]{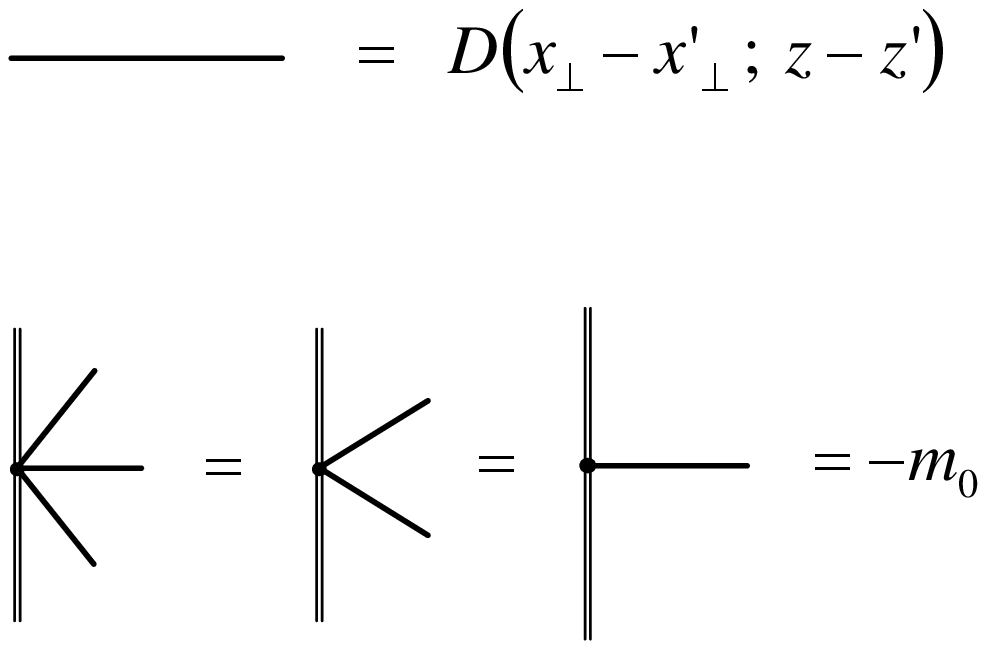}
\caption[]{Feynman rules obtained from the expansion of
(\ref{eff-Lagrangian})} \label{Feynman_rules_Sch}
\end{figure}

\begin{figure} [t!] \centering \noindent
\includegraphics[width=7cm]{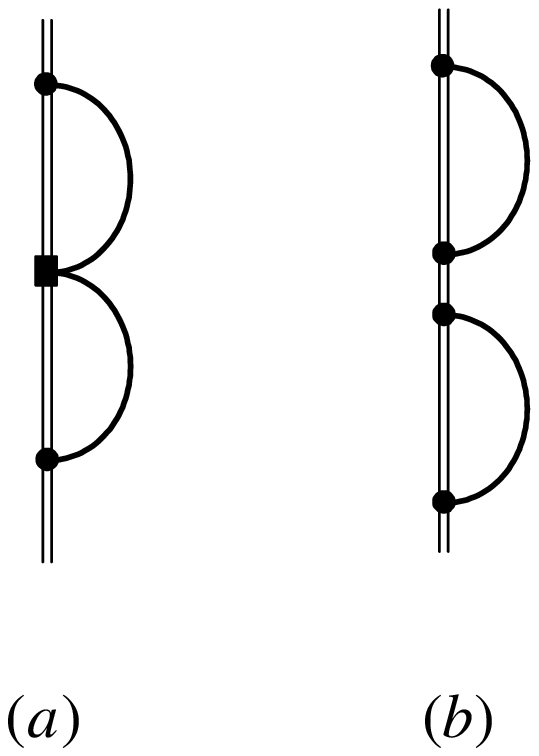}
\caption[]{(a) The leading finite-size contribution to $m$, which
is due to the term ${\cal O}$. The thick square vertex denotes an
insertion of ${\cal O}$. (b) A diagram that becomes disconnected
by the removal of the particle world-line. Therefore it does not
contribute to the computation of $m$ through the effective
action.} \label{Feynman_diag_Sch2}
\end{figure}

We turn to construct the Feynman rules, summarized in fig.\ref{Feynman_rules_Sch}.
Solid internal lines
denote the propagator for the scalar field $\phi$ on flat $\IR^{d-2,1} \times \IS^1$
\begin{equation}
D(x-x';z-z') = \frac{8 \pi G}{L} \frac{d-3}{d-2}
 \sum^\infty_{n=-\infty} \int \frac{d^{d-2} k_\perp}{(2\pi)^{d-2}}
 \frac{1}{k_\perp^2 + (2\pi n/L)^2}
 e^{i  k_\perp \cdot (x-x')_\perp +  2 \pi i n(z-z')/L},\label{scalar_prop}
\end{equation}
 where $k_\perp \equiv {\bf k}$.
The double solid line denotes the black hole world-line. There are
no propagators associated with this line. Finally the vertices
are constructed from the expansion of (\ref{eff-Lagrangian}) about
flat space. Those relevant for our computations
are listed on fig.\ref{Feynman_rules_Sch}. The Feynman rules for the fields in
 the decomposition (\ref{decompose}) are directly related to those of fig.\ref{Feynman_rules_Sch}.
 As usual, diagrams that become
disconnected by the removal of the particle world-line, such as
fig.\ref{Feynman_diag_Sch2}(b), do not contribute to the terms in
$\Gamma_{eff}[\bar{\phi},\overline{\delta\gamma}_{ij}]$.

\presub \emph{Power counting}. Each Feynman diagram in the ClEFT
contributes a definite power of $\lambda \sim \(\frac{r_0}{L}\)^{d-3}$
to the terms in (\ref{Gamma_eff}) and we now explain
how to evaluate this power for each diagram in a straightforward
manner (similar to \cite{CGR}). Our problem contains two dimensionful parameters
 $m_0=r_0^{d-3}$ and $L$. Powers of $r_0$ can come only from world-line vertices and not from
the bulk. Actually since we neglect finite-size effects each diagram is simply proportional to $m_0^{n_V}$ where $n_V$ is the number of world-line vertices. Powers of $L$ must arrange themselves automatically by dimensional analysis.

In more detail, since the only
scale in the propagator is $L$ (we set $G=c=1$) we assign $k\sim \del_i \sim
L^{-1}$ and thus \be
 D\sim\int{dk^{d-1} \over k^2} \sim L^{3-d}. \ee
 Based on this reasoning the propagators of vector and metric fields
$A_{i},\delta\gamma_{ij}$ are assigned the same scaling,
$L^{3-d}$. No scaling factors are assigned to the asymptotic
fields.

Altogether the diagrams of fig.\ref{Feynman_diag_Sch} scale like \bea
fig.\ref{Feynman_diag_Sch}(a) &\sim& m_0\lambda \non
fig.\ref{Feynman_diag_Sch}(b) &\sim& m_0\lambda^2 \eea
 and thus their contribution to $m$ is suppressed by a single and quadratic power of $\lambda$
respectively.

In order to count powers of finite-size higher-order terms in the
effective world-line action (\ref{eff-Lagrangian}) we note that by
dimensional analysis the dimension of a term sets the dimension of
its coefficient and $r_0$ is the only dimensionful parameter
which can enter into the expression for such a coefficients. For
example, the coefficient of $\co$ defined in (\ref{def-co}) must
be proportional to $m_0\, r_0^4 \propto r_0^{d+1}$ and the proportionality constant is
fixed by matching the effective Lagrangian of equation
(\ref{eff-Lagrangian}) to the full black hole theory, so that
observables calculated in the ClEFT agree with those of the full
theory.

The first finite-size correction to $m$ (through the constant term
in $\Gamma_{eff}[\bar{\phi},\overline{\delta\gamma}_{ij}]$) is due
to an insertion of $\co$ as in fig. \ref{Feynman_diag_Sch2}(a).
According to the power counting rules \bea
 {\cal O} &\sim& m_0\lambda^{4 \over d-3} \non
 fig. \ref{Feynman_diag_Sch2}(a) &\sim& m_0\lambda^{2(d-1) \over
d-3} \ll m_0\lambda^{2}\eea
 Therefore, the contribution of ${\cal O}$ along with other
finite-size higher derivative terms is always beyond second order
in $\lambda$, whereas for $d=5,6$ finite-size effects are beyond
third order.

\subsection{The renormalized mass at one-loop and higher}
 \label{m-subsection}

According to the definition of the perturbative regime (\ref{scales}), the small
parameter $\lam$ should be proportional to a (positive) power of $r_0/L$.
We set the normalization of the small parameter to be \be
 \lambda := \(\frac{r_0}{L}\)^{d-3}\, \zeta(d-3)  =
 \frac{16 \pi\, G\, m_0}{(d-2) \Omega_{d-2}\, L^{d-3}}\, \zeta(d-3)
\label{def-lambda} \ee
 where the equality relates the Schwarzschild radius $r_0$ in the
first expression to the  mass $m_0$ in the second expression and $\Omega_{d-2}:=(d-1) \pi^{(d-1)/2}/\Gamma[(d+1)/2]$
is the area if $\IS^{d-2}$. This normalization is such that $\phi(O)-=-\lambda + \dots$
as can be seen either from of the \Schw metric or explicitly in (\ref{phi-R-static}).
Hence $\lambda$ could be interpreted as the (absolute value of the)
Newtonian potential (to leading order). \footnote{This definition contains an extra factor
of $(d-3)/(d-2)$ relative to propagator of a canonically normalized scalar field which originates from
the pre-factor of the kinetic term for $\phi$ in the action (\ref{bulk_action}). If we were to compute the Newtonian potential in the original action, prior to dimensional reduction, this same factor would have emerged from the graviton propagator in the standard Feynman gauge. In this sense we get insight to this pre-factor in the action which is somewhat curious at first sight.}

The first correction to the mass of the system arises from the
1-loop diagram of fig.\ref{Feynman_diag_Sch}(a). Using the Feynman
rules of the ClEFT (see fig.\ref{Feynman_rules_Sch}) this diagram
is evaluated to be \be
 fig.\ref{Feynman_diag_Sch}(a)
  = \frac{\lam}{2}\, m_0 ~.
 \label{1-loop_mass} \ee
 This reproduces the results of \cite{H4},\cite{dialogue2},\cite{CGR}. It can be understood in Newtonian terms by comparing to the expression of the total Newtonian gravitational energy $E=\int \phi\, dm/2$ (like in electro-statics).

Appendix \ref{AppendixA} contains details of the derivation or
(\ref{1-loop_mass}) . Basically the loop gives the factor of
$\lam$ while the $1/2$ is a symmetry factor. An interesting  point
is that from the perspective of the wave-number space the sum over
the Kaluza-Klein harmonics gives a factor of $\zeta(4-d)$ while
from the configuration space perspective we expect the Newtonian
potential to be proportional to the sum $\sum_n 1/(n L)^{d-3}$
which is proportional to $\zeta(d-3)$. It turns out that the two
can be traded according to an identity involving the functions
zeta and gamma.

The regularization is a second noteworthy point about the derivation. In appendix \ref{AppendixA} we use dimensional regularization, while within the method of MAE \cite{dialogue2} advocated Hadamard's regularization which was claimed to be equivalent to omitting self-interaction terms (``no-SI''). Since both dimensional regularization and Hadamard's are essentially analytic continuations they are guaranteed to agree, but in this case we can moreover see explicitly the equivalence with no-SI. Considering the sum over $n$, the quantized KK wavenumber, the only divergent term is the one with $n=0$, while $n \neq 0$ can be thought to arise from the images of the black hole (in the covering space). Dimensional regularization puts the $n=0$ term to zero which indeed amounts to omitting self-interaction, keeping only the interaction with the images.

\presub \emph{2-loop}. The next contribution to $m$  is
suppressed by a factor of $\lambda$ relative to the 1-loop result and is given by a 2-loop diagram in
fig.\ref{Feynman_diag_Sch}(b). Using the same Feynman rules as before
we obtain \be
 fig.\ref{Feynman_diag_Sch}(b)=-{m_0 \over 2}\, \lambda^2 ~.
 \ee
Again, the factor $1/2$ is a symmetry factor. Adding up, we reproduce the
 result of \cite{CGR} up to second order in $\lambda$  \be m=m_0
\(1-{1 \over 2}\lambda+{1 \over 2}\lambda^2+\dots\)
\label{Sch_mass} \ee

\begin{figure} [t!] \centering \noindent
\includegraphics[width=7cm,angle=+90]{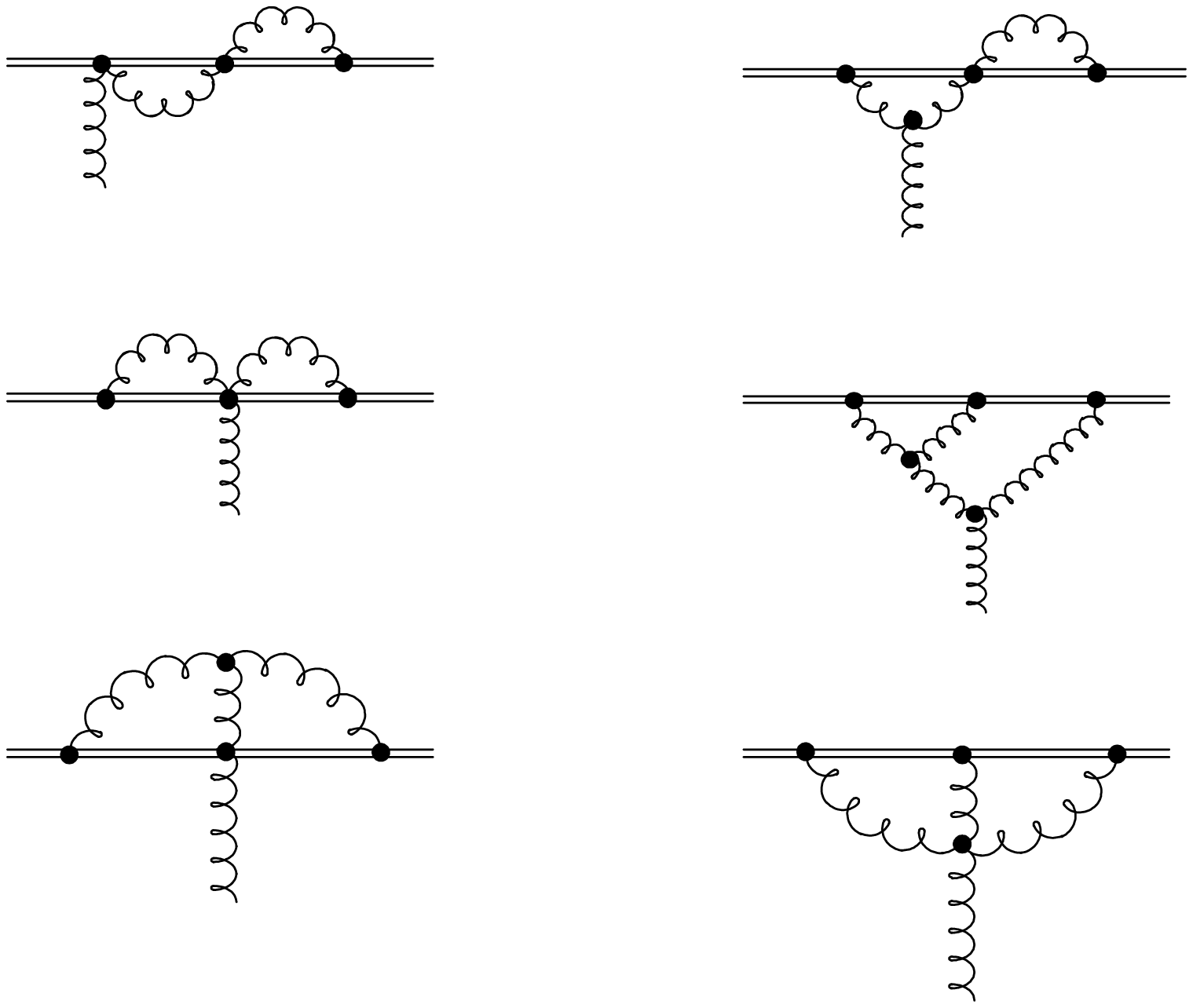}
\caption[]{The six 2-loops diagrams which were computed in
\cite{CGR} to determine $m$ to order $\co\(\lambda^2\)$. Compare
this with the single diagram fig.\ref{Feynman_diag_Sch}(b) which
is required by our improved method.} \label{sixdiag}
\end{figure}

Note that, whereas \cite{CGR}
computed six 2-loop diagrams (fig.\ref{sixdiag}) each with one
external leg, including a diagram with the quartic coupling of
GR, we compute a single 2-loop diagram with no external legs.
Moreover our diagram happens to factorize into two integrals,
which explains the factorization observed by \cite{CGR} for the
sum of their diagrams.

\begin{figure} [t!] \centering \noindent
\includegraphics[width=10cm] {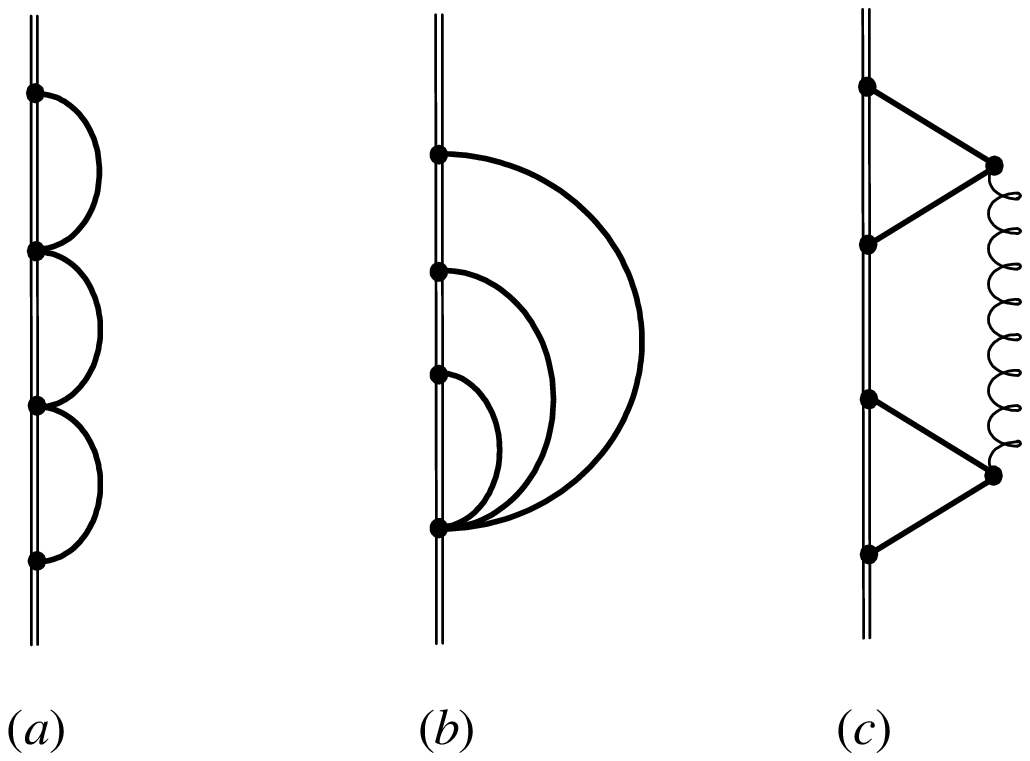}
\caption[]{Corrections to the mass $m$ of order
$\lambda^3$.} \label{Feynman_diag_Sch4}
\end{figure}

\presub \emph{Higher order corrections for $d=5,6$}. According to
the power counting rules established in subsection
\ref{Action&Feynman_rules}, finite-size effects do not contribute to $m$ at
order ${\cal O}(\lambda^3)$ for $d=5,6$. The relevant diagrams are those of
fig. \ref{Feynman_diag_Sch4}. Using the Feynman rules of fig.\ref{Feynman_rules_Sch} yields \bea
 fig. \ref{Feynman_diag_Sch4}(a)={m_0 \over 2}\lambda^3 \non
 fig. \ref{Feynman_diag_Sch4}(b)={m_0 \over 6}\lambda^3 ~.\eea
 Combining altogether we obtain for the ADM mass in $d=5,6$ up to the evaluation of the non-factorizable diagram  fig.\ref{Feynman_diag_Sch4}(c)
%3-loop
\be
 m=m_0\(1-{1 \over 2}\lambda+{1 \over 2}\lambda^2-\( {2 \over 3} +\frac{1}{m_0\, \lam^3}\, fig.\ref{Feynman_diag_Sch4}(c) \)\lambda^3+\dots\)
  \label{m-caged-3-loop} \ee

\subsection{Thermodynamics}

In order to calculate other thermodynamic quantities including the
tension and the entropy, we find it convenient to use the free
energy potential which, as explained in section \ref{thermo}, plays a
fundamental role in the system under consideration.

We start by calculating the red-shift factor (\ref{red-shift}). For
this purpose we use the effective action (\ref{eff-Lagrangian}) in
order to calculate the value of $\phi(O)$ at the black hole
location. Up to the second order in $\lambda$ the diagrams
contributing to the red-shift appear on fig.\ref{g00}.

Their value is given by \bea
 fig.\ref{g00}(a)&=& -\lambda \non
 fig.\ref{g00}(b)&=& \lambda^2\eea
As a result we obtain \bea
  \phi(O) &=& -\lambda+\lambda^2+\dots \non
   R &=& \sqrt{g_{00}(O)}=e^\phi=1-\lambda+{3 \over
 2}\lambda^2+\dots \label{phi-R-static} \eea
  Altogether the asymptotic temperature is given by
 \be
 T = R\, T_0
 = T_0\(1-\lambda+{3 \over 2}\lambda^2 +\dots\) \label{Sch_temperature} \ee
 where $T_0={ d-3 \over 4\pi r_0}$ is the local temperature of the black hole.

Next we relate the free energy $F=m-T S$ to the asymptotic charges
$m,\, \htau$ using Smarr's relation $(d-3)m=(d-2)T\, S+\htau\, L$, where $S,\, \htau$ are the entropy and the tension of the black hole
 respectively \footnotemark[3] % refers to line 835 roughly
 to eliminate the term $T\, S$
 \be
  m=(d-2)F- \htau L ~. \ee
 In this equation $m$ is known from (\ref{Sch_mass}) while $F$ is unknown. Considering the tension to
be a derivative of $F$ via $\htau=\del F/\del L$ we get a
differential equation which we can solve for $F$.
 Since we use $(m_0,\,L)$ as our basic variables,
we need to express $\htau$ accordingly\footnote{These are nothing but the standard relations for changing variables, in thermodynamics or otherwise, stated concisely \`{a} la Landau-Lifshitz \cite{LL}
in terms of Jacobians.
 We use ${\partial (u,v) \over \partial (x,y)}$ to denote the Jacobian determinant \be {\partial (u,v) \over
  \partial (x,y)}= \det \left(
                    \begin{array}{cc}
                      \frac{\partial u}{\partial x} & \frac{\partial u}{\partial y} \\
                      \frac{\partial v}{\partial x} & \frac{\partial v}{\partial y} \\
                    \end{array}
                  \right)
  \ee}
   \be
    \htau=\( {\partial F \over \partial L}\)_{T}
  = {\partial(F,T) \over \partial (m_0,L)}{\partial(m_0,L) \over \partial (L,T)}
 = -{\partial(F,T) \over \partial (m_0,L)}\, \({\partial T \over \partial m_0}\)_{L}^{-1}
  \ee
  Substituting n the expression for $T$ (\ref{Sch_temperature}) and solving the resulting
differential equation order by order in
  $\lambda$ we obtain \be F(m_0,L)={m_0 \over d-2}\(1+{ d-4 \over 2}\lambda+{7-2d \over
  2}\lambda^2+\dots\)\ee Having the expression for the free energy
  at hand one can compute all the thermodynamic quantities of
  interest. We list them below \bea
   {{\htau L\over m_0}} &=& {1 \over 2}(d-3)\lambda - (d-3)
  \lambda^2+\dots\non S &=& -\({\partial F \over \partial T}\)_{L}=-\({\partial F \over \partial m}\)_{L}\({\partial T \over \partial m}\)_{L}^{-1}
  =   S_0\, (1+0\cdot\lambda+0\cdot\lambda^2\dots) \label{Sch_tension_entropy} \eea
 where $S_0=\Omega_{d-2}\, r_0^{d-2}/(4 G)$ is the entropy of an uncompactified black hole.
The expression for the tension is identical to the corresponding
one in \cite{CGR}, whereas the entropy at first sight looks
different, but turns out to agree. To see that one needs to express the local mass $m_0$ in terms
of the asymptotic one $m$ through the relation (\ref{Sch_mass}),
substitute it in (\ref{Sch_tension_entropy}) and expand the result
in powers of $\lambda$. Our result merely states that entropy gets
no corrections up to a second order in $\lambda$, though we may
expect it to change when finite-size effects are taken into
account and the black hole is seen to deform from spherical
symmetry.

 \subsection{Generalization to all $X$}
\label{arbit-subsection}

Consider generalizing the previous analysis from an $\IS^1$
compactification to a general compactification manifold $X$. One
observes that although some Feynman rules change the diagrams to
be computed are the same. Actually, there is no change in the
vertices for the effective world-line action of the black hole,
and the only change enters through the propagator.

Since all the results up to this order depend on a single quantity
$\lambda$ it is sufficient to generalize the definition of
$\lambda$ (\ref{def-lambda}), and to define it to be the value of the Newtonian
potential at the location of the black hole, or in formulae \be
 \lambda :=  \left| \phi(O) \right| \label{gen-lam} \ee
 where $\phi$ solves the linearized $\phi$ equation of motion (\ref{phiEOM-lin})
this time on $\IR^d \times X$ and $\phi(O)$ is the constant term in the
Laurent series for $\phi$ around the origin.

This is a definition of $\lambda$ through a linear partial
differential equation that in general may be solved through
numerical relaxation. In some cases an analytic solution may be
available such as in our case $X=\IS^1$ where the method of
images serves, as well as in the more general case of the
$n$-dimensional torus $X=\IT^n$.

Summarizing, our results (\ref{m-caged-3-loop},\ref{Sch_temperature},\ref{Sch_tension_entropy})
 %3-loop or \ref{Sch-mass}
generalize to an arbitrary
compactification manifold $X$ once the definition of $\lam$ (\ref{def-lambda}) is generalized to (\ref{gen-lam}).

\section{Application: rotating caged black hole}
 \label{rotating-section}

In this section we propose an extension of the ClEFT approach to
black hole thermodynamics which includes spin. We obtain the leading
spin vertices in the world-line action.
We compute the leading $O(\lambda)$ corrections to the
thermodynamic quantities $m,\htau,S$ and angular-momentum $J$.
We supplement the power counting rules of the previous section
with the scaling of the angular momentum, and proceed to compute
$m$ and $J$ to the next to leading order. Finite-size effects do
 not contribute at this order.

 \subsection{Action and Feynman rules}

We consider a stationary spinning black hole in the same
background as in the static case --- $\IR^{d-2,1} \times \IS^1$.
The local angular momentum tensor which is measured by an
``intermediate'' observer at a distance $r_0 \ll r \ll L$ from the
black hole is denoted by $J_0^{\mu\nu}=-J_0^{\nu\mu}$
(conventionally normalized such that in 4d $|J_{12}|=|J_3|$),
whereas we denote the asymptotic angular-momentum by $J_{\mu\nu}$.
The rest of the notation is left unchanged.

After compactifying the extra dimension one loses the rotational
symmetry between the compact and extended dimensions, therefore the
angular momenta associated with these rotations are no longer
conserved and should be set to zero in a stationary phase.
Actually any rotation in such a plane would ultimately dissipate
into gravitational waves due to the compactification-induced
quadrupole moment of the black hole which
would create a varying quadrupole moment once rotation starts.
Temporal components of the angular momentum tensor vanish
as well by its definition as the momentum conjugate to rotations.
Combining altogether yields \be
J_0^{0\mu}=J_0^{z\mu}=0 \ee
 Therefore in general our system is characterized by  $\[{d-2 \over 2}\]$
parameters $J_0^A$, where $\[{d-2 \over 2}\]$
 is the rank of $SO(d-2)$ (the dimension of the Cartan subalgebra). $J_0^A$ are then the
angular momenta associated with commuting rotations in the
corresponding planes.

Let us discuss the terms that need to be added to the black hole effective action.
Integrating out short degrees of freedom $g_S$ replaces the
space-time in the vicinity of the horizon with an effective
Lagrangian for the black hole world-line coupled to gravity. In
general, such an effective action includes an infinite set of
possible non-minimal couplings of the point object to the
space-time metric.
The mass term, a universal part of the action which is independent of the
object's structure,
is given by (\ref{Sch_world_line_action}) for a spinless particle
(static black hole) and needs to be supplemented in our case by
including the spin degrees of freedom. The procedure for constructing
the action for the spinning point particle can be found in
\cite{ReggeHanson,Porto:2006bt,Porto:2005ac},
and here we are satisfied with mentioning some key points. The
rotation degrees of freedom (of a rigid body) are represented on the
world-line by a frame variable $e^\mu_I(t)$, where $I$ is a
``body'' index while $\mu$ is a space-time index. The angular
velocity is defined to be $\Omega^\mu_\nu:=e^I_\nu\, D_\tau
e^\mu_I$ where $D_\tau \equiv \dot{X}^\rho\, D_\rho$  is a
covariant derivative in the direction tangent to the world-line.
Due to the isotropy of the object, the action depends on $e^\mu_I$
only through $\Omega^\mu_\nu$ and actually the terms of interest
to us can be obtained from the term $\half\, I\,
\Omega^2 \subset S$, where the inertia tensor $I$ is related to the angular
momentum via $J_{\mu\nu}=I \Omega_{\mu\nu}$.

The leading order terms in $S_{BH}$ which involve $J_0$ are \be
 S_{SG} =
  {1 \over 2}\int J_0^{\,\alpha\beta}h_{\alpha\gamma,\beta}\, \dot x^{\gamma}dt+{1
\over 4}\int J_0^{\beta\gamma}\({1\over 2}h_{\beta\lambda,\mu}+
h_{\mu\lambda,\beta}-h_{\mu\beta,\lambda}\)h_\gamma^\lambda \,\dot
x^{\mu} dt+\dots, \ee
 where $``SG"$ stands for spin-gravity interaction; the metric perturbation $h_{\mu\nu}$
is defined by $g_{\mu\nu}=\eta_{\mu\nu}+h_{\mu\nu}$ where $g_{\mu\nu}$ is the metric prior to
the dimensional reduction (\ref{KKansatz}); in the perturbative regime (\ref{scales}) $h_{\mu\nu}$ can be considered to be small;
and the ellipsis denote terms which are of higher order in $h$ (and proportional to $J_0$). Combining with (\ref{bulk_action}) and (\ref{Sch_world_line_action}) yields  \bea
 S_{eff}[\phi,A_i,\gamma_{ij}] &=& -\frac{1}{16\pi G} \int dx^{d-1} \sqrt{\gamma}
  \[ R[\gamma] + \frac{d-2}{d-3} \(\del \phi\)^2 - \frac{1}{4}\, e^{2(d-2)\phi/(d-3)}\, F^2
  \] \non
   &-& m_0 -m_0\, \phi(O)-{m_0 \over 2}\, \phi(O)^2 +{J_0^{ij}\over 2}\, F_{ij}(O)\(\half+\frac{d-2}{d-3}\phi(O)\) \non
  &-& {J_0^{ij} \over 2}\, A_{i}(O)\, \del_j\phi(O) - \frac{J_0^{ij}}{4}\, \delta\gamma_{j}^{~~k}(O)\,  F_{ik}(O) + \dots \label{Kerr_eff_action}\eea

In this action we decompose the metric
tensor, the vector field and the scalar field into a long wavelength
non-dynamical background fields $\bar
\phi,\, \bar A_i$ and $\bar \gamma_{ij}$ which live at the asymptotic region and the short wavelength fields
$\phi_L,\, A_{Li},\, \gamma_{Lij}$  which include the scales of order $L$ \bea
 \phi &=& \phi_L+\bar \phi \non
 A_{i} &=& A_{Li}+\bar A_i \non
 \gamma_{ij} &=& \gamma_{Lij} + \bar{\gamma}_{ij}= \gamma_{Lij} + \eta_{ij}+\overline{\delta\gamma}_{ij}
 \eea

\begin{figure} [t!] \centering \noindent
\includegraphics[width=5cm,angle=-90] {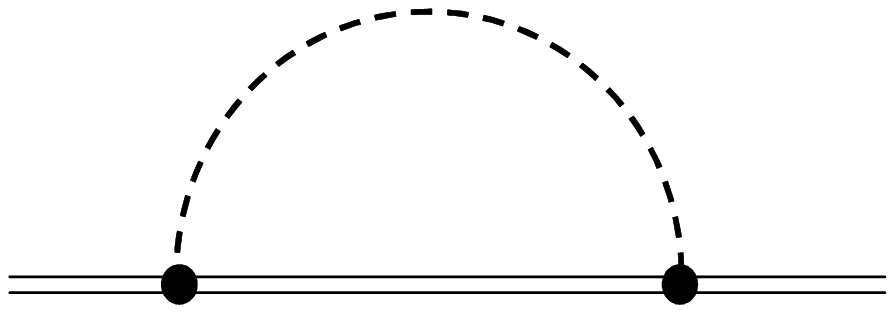}
\caption[]{A diagram which represents the spin-spin contribution to the renormalized mass $m$ according to the effective action (\ref{Gamma_eff_Kerr}). It is of order $J_0^2 \sim m_0\, \lambda^{d-1 \over d-3}$.} \label{Feynman_diag_Kerr1}
\end{figure}

\begin{figure} [t!] \centering \noindent
\includegraphics[width=12cm] {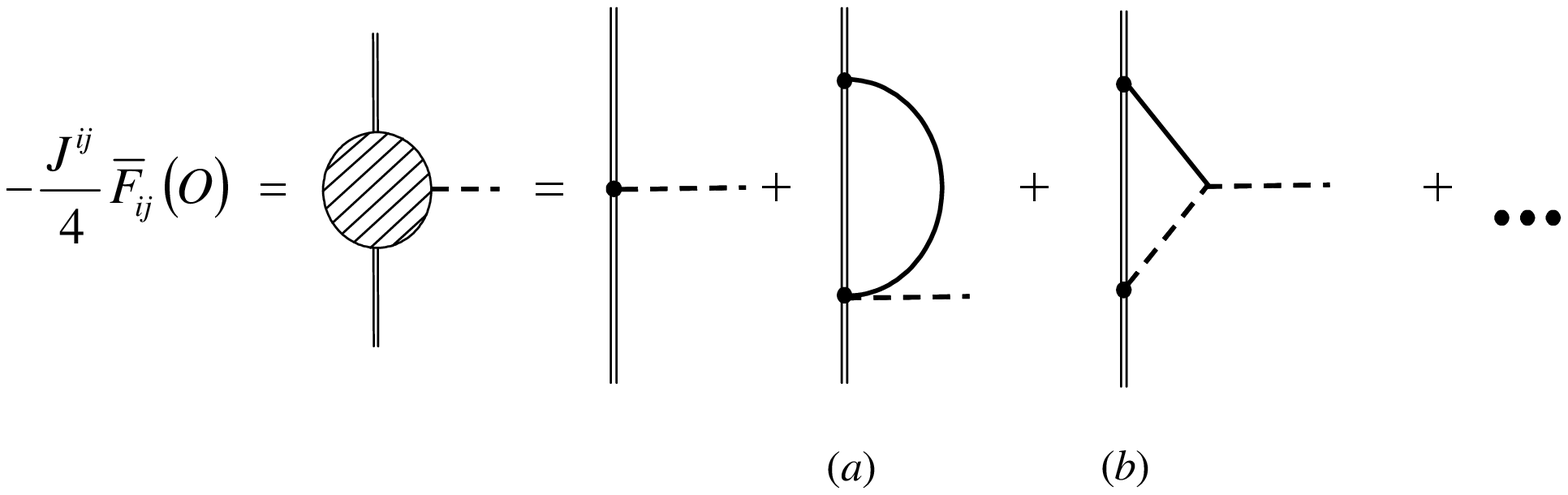}
\caption[]{A diagrammatic representation of the definition of the renormalized angular momentum $J$ according to the effective action (\ref{Gamma_eff_Kerr}). Both (a) and (b) represent corrections of order $\lambda$. }
\label{Feynman_diag_Kerr2}
\end{figure}

We now define the renormalized mass $m$ and angular momentum $J$. Integrating out in the sense of (\ref{Cl-int-out}) or equivalently
(\ref{FPI}) the short wavelength fields $\phi_L, A_{Li},\, \gamma_{Lij}$,
while holding the black hole world-line fixed leads to an
effective action $\Gamma_{eff}[\bar\phi,\bar A,\overline{\delta\gamma}]$ valid on
a scale much larger than $L$. The relation between the ADM mass
$m$ for a rotating caged black hole and the local mass $m_0$ along
with the relation between the local angular-momentum tensor
$J_0^{ij}$ and the asymptotic one $J^{ij}$ can be read off  \be
 \Gamma_{eff}[\bar\phi,\bar A,\overline{\delta\gamma}] = -m -m\, \bar\phi(O)+{J^{ij}\over 4}\bar F_{ij}(O)+\dots \label{Gamma_eff_Kerr}\ee
 The mass $m$ is the sum of Feynman diagrams like those of
fig.\ref{Feynman_diag_Sch} and fig.\ref{Feynman_diag_Kerr1},
whereas $J$  is given by the sum of tadpole diagrams
like those of fig.\ref{Feynman_diag_Kerr2}. As always, diagrams that become disconnected by the removal of the particle world-line do not contribute to the effective action $\Gamma_{eff}[\bar\phi,\bar A,\overline{\delta\gamma}]$.

\begin{figure}[t!]
\centering \noindent
\includegraphics[width=9cm]{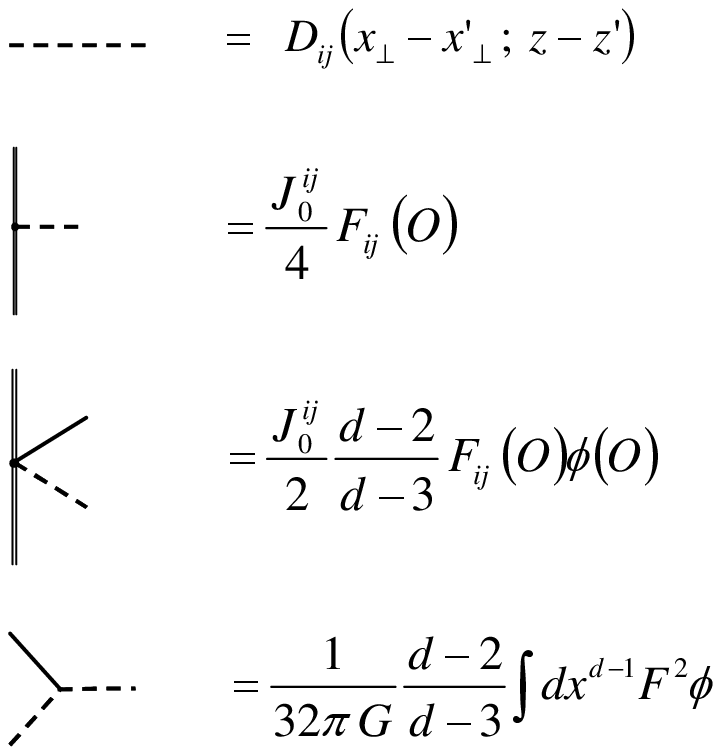}
\caption[]{Feynman rules obtained from the expansion of
(\ref{Kerr_eff_action})} \label{Feynman_rules_Kerr}
\end{figure}

The additional Feynman rules beyond those of fig.\ref{Feynman_rules_Sch} are summarized in fig.\ref{Feynman_rules_Kerr}. The dashed lines
denote the propagator for the vector field $A_i$ on flat
$\IR^{d-2,1} \times \IS^1$
\begin{equation}
D_{ij}(x-x';z-z') = -{16 \pi G \over L} \sum^\infty_{n=-\infty}
\int {d^{d-2} k_\perp \over (2\pi)^{d-2}} {\delta_{ij} \over
k_\perp^2 + (2\pi n/L)^2} e^{i k_\perp \cdot (x-x')_\perp +  2 \pi
i n(z-z')/L},\label{vector_prop}
\end{equation}
 where we used the Feynman gauge defined by adding to the action (\ref{Kerr_eff_action}) the following gauge fixing term \be
 S_{GF}={1 \over 32\pi G}\int d^{d-1}x\, (\del^i A_i)^2
\ee
 The vertices in the bulk are constructed from the
expansion of (\ref{Kerr_eff_action}) about flat space. Those relevant for our computations are summarized in
fig.\ref{Feynman_rules_Sch} and fig.\ref{Feynman_rules_Kerr}.

Counting powers of $r_0$ needs to be supplemented by a scaling of
 $J_0$ and we assign $J_0 \sim m_0 \, r_0$.

 \subsection{The renormalized mass and angular momentum}

Here we calculate the renormalized mass according to fig.\ref{Feynman_diag_Sch}. The leading $O(\lambda)$ correction to $m$ is still nothing but the 1-loop diagram in fig.\ref{Feynman_diag_Sch}(a).
Therefore the leading order contribution to the mass of the rotating
black hole is identical to the static case and is given by
(\ref{1-loop_mass})
 \be m=m_0 \(1-{1 \over 2}\lambda+\dots \)
\label{Kerr_1-loop_mass} \ee

In order to calculate the leading contribution
(of order $m_0\, J_0$) to the angular-momentum $J^{ij}$ one needs to compute the 1-loop
tadpole diagrams of fig.\ref{Feynman_diag_Kerr2}(a),(b). Using the
Feynman rules listed on fig.\ref{Feynman_rules_Sch} and
fig.\ref{Feynman_rules_Kerr} we obtain \bea
 fig.\ref{Feynman_diag_Kerr2}(a)&=& -{d-2 \over d-3}\lambda {J_0^{ij} \over 2}\bar F_{ij}(O)\non
 fig.\ref{Feynman_diag_Kerr2}(b)&=& {d-2 \over d-3}\lambda {J_0^{ij} \over 2}\bar F_{ij}(O) \label{4.9}
 \eea
As a result, the overall contribution to $J^{ij}$ vanishes at
linear order in $\lambda$ \be
 J^{ij}=J_0^{ij}(1+0\cdot\lambda+\dots )\label{ang_moment} ~.\ee
 Actually, there can be no contribution to $J$ of order $J_0^2$ either due to the absence
 of a cubic vertex for the vector field.

\presub \emph{Higher order correction}.
Fig.\ref{Feynman_diag_Kerr1} represents a Feynman diagram which
contributes to the mass $m$ at the next to leading order
$O(\lambda^{1+{2\over d-3}})$. Applying the Feynman rules of
fig.\ref{Feynman_rules_Kerr} it is evaluated to be \be
 fig.\ref{Feynman_diag_Kerr1} =
  -{d-2 \over 4}{J^{ij}J_{ij}\over m_0 r_0^2}\( {r_0 \over L} \)^{d-1} \zeta(d-1) \label{jj-diag} \ee
(see Appendix \ref{AppendixA} for details). As a result, the mass of
the rotating caged black hole (\ref{Kerr_1-loop_mass}) is modified
\be
 m = m_0 \(1-{1 \over 2}\lambda+{d-2 \over 4}{J_0^{ij} J_{0ij}\over
(m_0 r_0)^2}\( {r_0 \over L} \)^{d-1}\zeta(d-1)+\dots \)
 \label{m-rot} \ee
 We note that this result is consistent with the 4d spin-spin interaction \\
 $V_{SS}=+G\(3 (\vec{S}_1 \cdot \hat{r})\, (\vec{S}_2 \cdot \hat{r})-\vec{S}_1 \cdot \vec{S}_2 \)/r^3$ \cite{Wald-spin}, see also \cite{Porto:2005ac}.

\subsection{Thermodynamics}

In this section we derive additional thermodynamic quantities
through the use of the Gibbs potential. We consider only the leading order
corrections to the thermodynamic quantities. For simplicity
and without loss of generality we assume that only one of the
local spin parameters $J_0^A$ is non-zero and we denote it by $J_0$.

The calculation of the red-shift factor up to linear order in
$\lambda$ does not differ from the static case. Thus the temperature
and the angular velocity possess the same form as
(\ref{Sch_temperature}) \bea
  T &=& R\, T_0=T_0\(1-\lambda+\dots \) \non
 \Omega &=& R\, \Omega_0=\Omega_0\(1-\lambda+\dots \)
 \label{Kerr_temperature} \eea
 where $T_0$,$\Omega_0$ are the local
temperature and angular velocity of the rotating Myers-Perry
 black hole \cite{MyersPerry} \bea
 T_0 &=& {d-5 \over 4\pi r_+} +{1 \over 2\pi}{r_+^{d-4} \over r_0^{d-3}} \non
\Omega_0&=&{a \over r_+^2+a^2} \eea
 $a$ denotes the rotation parameter in terms of which the hole's
 angular momentum is \be
 J_0 = {2 m_0 a \over (d-2)} ~, \label{def-a} \ee
 and $r_+$ is the location of the horizon given implicitly by \be
  r_0^{d-3} = r_+^{d-5}(r_+^2+a^2) ~.\ee
For later use we list also the hole's entropy \be
  S_0 = {\pi^{d-3 \over 2} \over 4\,GT_0\Gamma({d-3 \over 2})}r_0^{d-3}\(1-{2 \over d-3}{a^2
\over r_+^2+a^2}\) = \frac{4 \pi\, r_+\, m_0}{d-2} ~, \label{def-S} \ee
 where only the first expression appears in \cite{MyersPerry}.

We now wish to translate our knowledge of the thermodynamic potential
$m$ into the Gibbs potential $G$ which is more appropriate for the natural variables of the problem.
Using the definition of the Gibbs potential (\ref{Gibbs})
and the Smarr formula $(d-3)m=(d-2)(\Omega J+TS)+\htau L$ yields the simple relation
\be m=(d-2)G-\htau L ~.\label{Smarr_Gibbs}  \ee
 In this equation $m$ is known from (\ref{Kerr_1-loop_mass}), while $G$ is unknown.
Considering the tension to be a derivative of $G$ via $\htau=\del G/\del L$ we get a
 differential equation which we can solve for $G$.

As mentioned in subsection \ref{thermo}, the natural parameters for the computation are $a$ and $m_0$
rather than $\Omega$ and $T$, therefore one needs to establish the
relations between the derivatives of the Gibbs potential expressed
in terms of these two sets. We list some useful relations\footnotemark[8] %\cite{LL} around line 1407
 \bea
 \htau &=& \( {\partial G \over \partial L}\)_{T,\,\Omega}
  = {\partial(G,T,\Omega) \over \partial (L,T,\Omega)}={\partial (G,T,\Omega) \over \partial (L,m_0,a)}\({\partial(L,T,\Omega) \over \partial
  (L,m_0,a)}\)^{-1}\non
J&=&-\( {\partial G \over \partial
\Omega}\)_{T,\,L}=-{\partial(G,T,L) \over \partial
    (\Omega,T,L)}=-{\partial(G,T,L) \over \partial
    (m_0,a,L)}\({\partial(\Omega,T,L) \over \partial
    (m_0,a,L)}\)^{-1}\non
S&=&-\( {\partial G \over \partial
T}\)_{\Omega,\,L}=-{\partial(G,\Omega,L) \over \partial
    (T,\Omega,L)}=-{\partial(G,\Omega,L) \over \partial
    (m_0,a,L)}\({\partial(T,\Omega,L) \over \partial
    (m_0,a,L)}\)^{-1} \label{Kerr_therm_quantities}
  \eea

The expression for $\htau$ simplifies at the leading order in $\lambda$ and we obtain \be
  \htau = \( {\partial G \over \partial L}\)_{T,\,\Omega}
  =\( {\partial G \over \partial L}\)_{m_0,\,a}+{m_0 \over L}{(d-3)^2 \over
  (d-2)}\lambda+\dots \label{tau}\ee
In appendix \ref{AppendixB} we present useful identities for the derivation
of the above relation for $\htau$ and those which follow.

Substituting (\ref{tau}) and (\ref{Kerr_1-loop_mass}) in the relation between $m$ and $G$
(\ref{Smarr_Gibbs}) and solving the resulting differential
equation for $G(m_0,a,L)$ up to linear order in $\lambda$ yields
\be G(m_0,L)={m_0 \over d-2}\(1+{ d-4 \over 2}\lambda+\dots \)
 \ee
Substituting this expression into
(\ref{Kerr_therm_quantities},\ref{tau}) we finally obtain \bea
 {\htau L \over m_0} &=& {d-3 \over 2}\lambda+ \dots \non
 J &=& J_0\, (1 + 0 \cdot \lambda +\dots) \non
 S &=& S_0\, (1 + 0 \cdot \lambda \dots ) \label{rot-tau-j-s} \eea
 where $J_0,~S_0$ are defined by (\ref{def-a},\ref{def-S}).
This result for $J$, obtained through a 0-point function, reproduces (\ref{ang_moment}) obtained through
 a 1-point function. It states that
the angular-momentum is left ``unrenormalized" in the leading order. This agreement can be
considered a consistency check for the Feynman rule of the $J_0\, F\, \phi$ vertex in  fig.\ref{Feynman_rules_Kerr}. The other results are consistent with the static case \ref{Sch_tension_entropy}.

\subsection*{Acknowledgements}

It is a pleasure to thank Shmuel Elitzur for a discussion.

This research is supported by The Israel Science Foundation grant no
607/05,  DIP grant H.52, EU grant MRTN-CT-2004-512194 and the
Einstein Center at the Hebrew University.

\appendix

\section{Calculations for Feynman diagrams} \label{AppendixA}

In this appendix we calculate certain integrals denoted by $I_0,\, I_1$ and defined below,
 which are useful for evaluating the Feynman diagrams (\ref{1-loop_mass},\ref{jj-diag}), respectively.
Both integrals are ultraviolet divergent and we use
dimensional regularization.

We start from  \be I_0(L):= {1\over 2
L}\sum^\infty_{n=-\infty}\int {d^{d-2} k_\perp\over
(2\pi)^{d-2}}{1\over k_\perp^2 + (2\pi n/L)^2} \ee Let us use the
dimensional regularization result \be
 \int{d^D k \over (2\pi)^D}{1 \over (k^2+\Delta)^m }= {1\over (4\pi)^{D/2}} {\Gamma\(m-{D\over 2}\)\over
 \Gamma(m)}\Delta^{{D\over2}-m}\ee with $D=d-2,m=1$ and $\Delta=\({2\pi n \over
 L}\)^2$, then the $n=0$ term in $I_0(L)$ vanishes and the rest yields\be
  I_0(L)={{\pi^{d-6\over 2}}\over 4L^{d-3}}~\zeta(4-d)\,\Gamma\({4-d\over
  2}\) \label{I0first} \ee
 Note that $\Gamma[(4-d)/2]$ has a pole for $d=4,6,8,\dots$, while $\zeta(4-d)$ has a zero
for exactly the same values of $d$.
 We can avoid this feature by using a relation between the Gamma function and the Riemann zeta
  function \be
   \Gamma \({s\over 2}\)\pi^{-s/2}\zeta(s)= \Gamma \({1-s\over
   2}\)\pi^{-(1-s)/2}\zeta(1-s)\label{Gamma_Zeta_rel}\ee
  from which we get \be
 \Gamma(\frac{4-d}{2})\, \zeta(4-d)  = \pi^{7/2-d}\, \Gamma(\frac{d-3}{2})\, \zeta(d-3)  =
  \frac{4}{d-3}\, \frac{\pi^{3-d/2}}{\Omega_{d-2}}\, \zeta(d-3)  ~.\ee
  Substituting back into (\ref{I0first}) we finally obtain \be
    I_0(L)
  = {\Gamma\left({d-3\over 2}\right)\over (4\pi)^{d-1\over 2}}
\left({2\over L}\right)^{d-3}\zeta(d-3) \label{I0} .\ee

We now turn to $I_1$ defined by \be
I_1(L):={2 \over L}
\sum^\infty_{n=-\infty}\int {d^{d-2} k_\perp\over
(2\pi)^{d-2}}{k_\perp^2\over k_\perp^2 + (2\pi n/L)^2}  \label{def-I_1}\ee
 First we use the following dimensional regularization result \be
 \int{d^D k \over (2\pi)^D}{k^2 \over (k^2+\Delta)^m }= {D/2\over (4\pi)^{D/2}} {\Gamma\(m-{D\over 2}-1\)\over
 \Gamma(m)}\Delta^{{D\over2}-m+1} \ee
 with the same $D=d-2,m=1$ and $\Delta=\({2\pi n \over
 L}\)^2$ as before. The $n=0$ term in $I_1(L)$ vanishes and the rest yields \be
  I_1(L)={2\pi^{d-2\over 2}\over L^{d-1}}~(d-2)\zeta(2-d)\,\Gamma\({2-d\over
  2}\)\ee
 Finally, applying relation (\ref{Gamma_Zeta_rel}) gives \be
    I_1(L)=(d-2)(d-3){\Gamma\left({d-3\over 2}\right)\over \pi^{d-1\over 2}}
{\zeta(d-1)\over L^{d-1}} ~. \label{I1} \ee

\section{Useful thermodynamic identities}
 \label{AppendixB}

In this appendix we consider a spinning black hole imbedded in an
uncompactified $d$-dimensional space-time \cite{MyersPerry}. We assume
that only one of the spin parameters is non-zero and present
different identities valid in this case. These identities are found
to be useful for the derivation of the thermodynamics of a rotating caged black hole
 considered in the text. The notation is
explained in the text.

We start from \be
 {\del (T_0, \Omega_0) \over \del (m_0,a)} = {1 \over
2m_0a}{\del T_0 \over \del a} \label{identity_1} \ee
This equation restates the thermodynamic identity
$J_0=-\del G/\del \Omega$ in the $(m_0,a)$ set of variables \bea
 J_0 &=& \frac{2m_0\, a}{d-2} \non
 \( \frac{\partial G_0}{\partial \Omega_0}\)_{T_0}
 &=& {\partial(G_0,T_0) \over \partial (\Omega_0,T_0)}
 = {\partial(G_0,T_0) \over \partial (m_0,a)}\, \({\partial(\Omega_0,T_0) \over \partial (m_0,a)}\)^{-1} \non
 &=& \frac{1}{d-2}\, \frac{\del T_0}{\del a}\, \({\partial(\Omega_0,T_0) \over \partial (m_0,a)}\)^{-1}
 \eea
 where in the last equality we used that $G_0=m_0/(d-2)$ according to (\ref{Smarr_Gibbs})
with no $\htau$ in the uncompactified case.

Another set of useful identities can be obtained after taking
account of scaling dimensions. Indeed, performing a scaling
transformation $L\rightarrow(1+\epsilon)L$ and recalling that
$m_0,a,T_0,\Omega_0$ have length dimensions $d-3,1,-1,-1$ respectively, we
get by expanding \be \left(
  \begin{array}{c}
    dT_0 \\
    d\Omega_0\\
  \end{array}
\right) = \left(
           \begin{array}{cc}
             {\del T_0\over \del m_0}~ & {\del T_0\over \del a} \\\\
             {\del \Omega_0\over \del m_0}~ & {\del \Omega_0\over \del a} \\
           \end{array}
         \right)\left(
                  \begin{array}{c}
                    dm_0 \\
                    da \\
                  \end{array}
                \right)
\ee to first order in $\epsilon$ \be \left(
  \begin{array}{c}
    -T_0 \\
    -\Omega_0\\
  \end{array}
\right) = \left(
           \begin{array}{cc}
             {\del T_0\over \del m_0}~ & {\del T_0\over \del a} \\\\
             {\del \Omega_0\over \del m_0}~ & {\del \Omega_0\over \del a} \\
           \end{array}
         \right)\left(
                  \begin{array}{c}
                    (d-3)\, m_0 \\
                    a \\
                  \end{array}
                \right)
\ee This expression can be inverted and rewritten as follows \be
\left(
      \begin{array}{c}
       (3-d)m_0 \\
       -a \\
      \end{array}
\right)=\({\del(T_0,\Omega_0)\over\del(m_0,a)}\)^{-1}
         \left(
           \begin{array}{cc}
             {\del \Omega_0\over \del a}~ & -{\del T_0\over \del a} \\\\
             -{\del \Omega_0\over \del m_0}~ & {\del T_0\over \del m_0} \\
           \end{array}
         \right)\left(
  \begin{array}{c}
    T_0 \\
    \Omega_0\\
  \end{array}
\right) \ee Combining this result with (\ref{identity_1}), we
finally obtain \bea
 {3-d \over 2a}~{\del T_0 \over \del a} &=& T_0{\del \Omega_0 \over \del a}-\Omega_0{\del T_0 \over \del
 a} \non
 {1 \over 2m_0}~{\del T_0 \over \del a}  &=& T_0{\del \Omega_0 \over \del m_0}-\Omega_0{\del T_0 \over \del
 m_0}
\eea


\begin{thebibliography}{99}

\bibitem{Damour}
T. Damour, {\em Gravitational Radiation and the Motion of Compact
Bodies}, In: N. Deruelle and T. Piran, {\em Gravitional
Radiation}, (North-Holland Publishing Company, 1983).

\bibitem{dialogue1}
D.~Gorbonos and B.~Kol, ``A dialogue of multipoles: Matched
asymptotic expansion for caged black holes,'' JHEP {\bf 0406}, 053
(2004) [arXiv:hep-th/0406002].
%%CITATION = HEP-TH 0406002;%%

\bibitem{dialogue2}
  D.~Gorbonos and B.~Kol,
  ``Matched asymptotic expansion for caged black holes: Regularization of  the
  post-Newtonian order,''
  Class.\ Quant.\ Grav.\  {\bf 22}, 3935 (2005)
  [arXiv:hep-th/0505009].
  %%CITATION = CQGRD,22,3935;%%

%%%% Goldberger and Rothstein
\bibitem{CGR}
  Y.~Z.~Chu, W.~D.~Goldberger and I.~Z.~Rothstein,
  ``Asymptotics of d-dimensional Kaluza-Klein black holes: Beyond the
  newtonian approximation,''
  JHEP {\bf 0603}, 013 (2006)
  [arXiv:hep-th/0602016].
  %%CITATION = JHEPA,0603,013;%%

\bibitem{GoldbergerRothstein1}
  W.~D.~Goldberger and I.~Z.~Rothstein,
  ``An effective field theory of gravity for extended objects,''
  Phys.\ Rev.\  D {\bf 73}, 104029 (2006)
  [arXiv:hep-th/0409156].
  %%CITATION = PHRVA,D73,104029;%%

\bibitem{GoldbergerRothstein2}
  W.~D.~Goldberger and I.~Z.~Rothstein,
  ``Dissipative effects in the world-line approach to black hole dynamics,''
  Phys.\ Rev.\  D {\bf 73}, 104030 (2006)
  [arXiv:hep-th/0511133].
  %%CITATION = PHRVA,D73,104030;%%

\bibitem{Goldberger-Lect}
  W.~D.~Goldberger,
  ``Les Houches lectures on effective field theories and gravitational
  radiation,''
  arXiv:hep-ph/0701129.
  %%CITATION = HEP-PH/0701129;%%

%%%% Black-hole black-string transition
\bibitem{TopChange}
  B.~Kol,
  ``Topology change in general relativity and the black-hole black-string
  transition,''
  JHEP {\bf 0510}, 049 (2005)
  [arXiv:hep-th/0206220].
  %%CITATION = JHEPA,0510,049;%%

\bibitem{review}
    B.~Kol,
    ``The phase transition between caged black holes and black strings: A
    review,''
  Phys.\ Rept.\  {\bf 422}, 119 (2006)
  [arXiv:hep-th/0411240].
  %%CITATION = PRPLC,422,119;%%


\bibitem{HOrev}
  T.~Harmark, V.~Niarchos and N.~A.~Obers,
  ``Instabilities of black strings and branes,''
  Class.\ Quant.\ Grav.\  {\bf 24}, R1 (2007)
  [arXiv:hep-th/0701022].
  %%CITATION = CQGRD,24,R1;%%


\bibitem{H4}
  T.~Harmark,
  ``Small black holes on cylinders,''
  Phys.\ Rev.\  D {\bf 69}, 104015 (2004)
  [arXiv:hep-th/0310259].
  %%CITATION = PHRVA,D69,104015;%%

\bibitem{KSSW1}
  D.~Karasik, C.~Sahabandu, P.~Suranyi and L.~C.~R.~Wijewardhana,
``Analytic approximation to 5 dimensional black holes with one compact
  dimension,''
  Phys.\ Rev.\  D {\bf 71}, 024024 (2005)
  [arXiv:hep-th/0410078].
  %%CITATION = PHRVA,D71,024024;%%



%%%% Caged numeric
\bibitem{KPS}
  B.~Kol, E.~Sorkin and T.~Piran,
  %``Caged black holes: Black holes in compactified spacetimes. I: Theory,''
  Phys.\ Rev.\  D {\bf 69}, 064031 (2004)
  [arXiv:hep-th/0309190];
  %%CITATION = PHRVA,D69,064031;%%
%\bibitem{KPS2}
%  E.~Sorkin, B.~Kol and T.~Piran,
  ``Caged black holes: Black holes in compactified spacetimes. II: 5d
  numerical implementation,''
  Phys.\ Rev.\  D {\bf 69}, 064032 (2004)
  [arXiv:hep-th/0310096].
  %%CITATION = PHRVA,D69,064032;%%

\bibitem{KudohWiseman6d}
  H.~Kudoh and T.~Wiseman,
  ``Properties of Kaluza-Klein black holes,''
  Prog.\ Theor.\ Phys.\  {\bf 111}, 475 (2004)
  [arXiv:hep-th/0310104];
  %%CITATION = PTPKA,111,475;%%
 % 6d
%\bibitem{Kudoh6d}
  H.~Kudoh,
  ``6-dimensional localized black holes: Numerical solutions,''
  Phys.\ Rev.\  D {\bf 69}, 104019 (2004)
  [Erratum-ibid.\  D {\bf 70}, 029901 (2004)]
  [arXiv:hep-th/0401229].
  %%CITATION = PHRVA,D69,104019;%%
% 6d

\bibitem{KudohTanakaNakamura}
  H.~Kudoh, T.~Tanaka and T.~Nakamura,
  ``Small localized black holes in braneworld: Formulation and numerical
  method,''
  Phys.\ Rev.\  D {\bf 68}, 024035 (2003)
  [arXiv:gr-qc/0301089].
  %%CITATION = PHRVA,D68,024035;%%

\bibitem{KudohWiseman5d}
  H.~Kudoh and T.~Wiseman,
  ``Connecting black holes and black strings,''
  Phys.\ Rev.\ Lett.\  {\bf 94}, 161102 (2005)
  [arXiv:hep-th/0409111].
  %%CITATION = PRLTA,94,161102;%%
% 5d and 6d


\bibitem{KantiRev}
  P.~Kanti,
  ``Black holes in theories with large extra dimensions: A review,''
  Int.\ J.\ Mod.\ Phys.\  A {\bf 19}, 4899 (2004)
  [arXiv:hep-ph/0402168].
  %%CITATION = IMPAE,A19,4899;%%

%% rotating
\bibitem{MyersPerry}
  R.~C.~Myers and M.~J.~Perry,
  ``Black Holes In Higher Dimensional Space-Times,''
  Annals Phys.\  {\bf 172}, 304 (1986).
  %%CITATION = APNYA,172,304;%%

\bibitem{KKR}
  B.~Kleihaus, J.~Kunz and E.~Radu,
  ``Rotating nonuniform black string solutions,''
  JHEP {\bf 0705}, 058 (2007)
  [arXiv:hep-th/0702053].
  %%CITATION = JHEPA,0705,058;%%

%%%% Effective action for spinning point particle
\bibitem{ReggeHanson}
  A.~J.~Hanson and T.~Regge,
  ``The Relativistic Spherical Top,''
  Annals Phys.\  {\bf 87}, 498 (1974).
  %%CITATION = APNYA,87,498;%%

\bibitem{Porto:2006bt}
  R.~A.~Porto and I.~Z.~Rothstein,
  ``The hyperfine Einstein-Infeld-Hoffmann potential,''
  Phys.\ Rev.\ Lett.\  {\bf 97}, 021101 (2006)
  [arXiv:gr-qc/0604099].
  %%CITATION = PRLTA,97,021101;%%

\bibitem{Porto:2005ac}
  R.~A.~Porto,
  ``Post-Newtonian corrections to the motion of spinning bodies in NRGR,''
  Phys.\ Rev.\  D {\bf 73}, 104031 (2006)
  [arXiv:gr-qc/0511061].
  %%CITATION = PHRVA,D73,104031;%%


\bibitem{AdS-renorm}
  M.~Henningson and K.~Skenderis,
  ``The holographic Weyl anomaly,''
  JHEP {\bf 9807}, 023 (1998)
  [arXiv:hep-th/9806087].
  %%CITATION = JHEPA,9807,023;%%
 V.~Balasubramanian and P.~Kraus,
  ``A stress tensor for anti-de Sitter gravity,''
  Commun.\ Math.\ Phys.\  {\bf 208}, 413 (1999)
  [arXiv:hep-th/9902121].
  %%CITATION = CMPHA,208,413;%%
  S.~de Haro, S.~N.~Solodukhin and K.~Skenderis,
  ``Holographic reconstruction of spacetime and renormalization in the  AdS/CFT
  correspondence,''
  Commun.\ Math.\ Phys.\  {\bf 217}, 595 (2001)
  [arXiv:hep-th/0002230].
  %%CITATION = CMPHA,217,595;%%
%\bibitem{Skenderis-Lect}
  K.~Skenderis,
  ``Lecture notes on holographic renormalization,''
  Class.\ Quant.\ Grav.\  {\bf 19}, 5849 (2002)
  [arXiv:hep-th/0209067].
  %%CITATION = CQGRD,19,5849;%%

\bibitem{Wald-spin}
  R.~Wald,
  ``Gravitational spin interaction,''
  Phys.\ Rev.\  D {\bf 6}, 406 (1972).
  %%CITATION = PHRVA,D6,406;%%

\bibitem{HIW}
  S.~Hollands, A.~Ishibashi and R.~M.~Wald,
  ``A higher dimensional stationary rotating black hole must be
  axisymmetric,''
  Commun.\ Math.\ Phys.\  {\bf 271}, 699 (2007)
  [arXiv:gr-qc/0605106].
  %%CITATION = CMPHA,271,699;%%

\bibitem{LL} L.~D.~Landau and E.~M.~Lifshitz,
 {\it Statistical Physics} Part 1,
 (Pergamon Press, 1980),  \S 16. %p.51.

\end{thebibliography}
\end{document}